\documentclass[%
 reprint,
 amsmath,amssymb,
 pra,
]{revtex4-2}

\usepackage{graphicx}
\usepackage{dcolumn}
\usepackage{bm}
\usepackage[normalem]{ulem}
\usepackage[dvipsnames]{xcolor}

\newcommand{\tr}{\mathrm{tr}}

\newcommand{\iden}{\mathbb{I}}

\newcommand{\MTO}{\mathrm{MTO}}

\newcommand{\colli}{\mathcal{C}}
\newcommand{\cone}{\mathbb{CO}}

\newtheorem{theorem}{Theorem}

\newtheorem{definition}{Definition}

\newcommand{\mE}{\Gamma}
\newcommand{\mH}{\mathcal{H}}
\newcommand{\mC}{\mathcal{C}}

\begin{document}


\title{Thermalization of finite complexity and its application to heat bath algorithmic cooling}

\author{Xueyuan Hu}
 \email{xyhu@sdu.edu.cn}
 \affiliation{School of Information Science and Engineering, Shandong University, Qingdao 266237, China}

\author{Valerio Scarani}%
\email{physv@nus.edu.sg}
\affiliation{Centre for Quantum Technologies, National University of Singapore, 3 Science Drive 2, Singapore 117543}
\affiliation{Department of Physics, National University of Singapore, 2 Science Drive 3, Singapore 117542}


\date{\today}

\begin{abstract}
We introduce a class of thermal operations based on the collision model, where the system sequentially interacts with uncorrelated bath molecules via energy-preserving unitaries. To ensure finite complexity, each molecule is constrained to be no larger than the system. We identify a necessary condition for cooling below the bath temperature via a single collision: the system must initially lack a well-defined effective temperature, even a negative one. By constructing a iterative protocol, we demonstrate that sub-bath cooling is achievable without a machine under these restricted thermal operations. Moreover, introducing a qubit machine further enhances both the cooling limit and energy efficiency. These findings contribute to the broader study of cooling with finite resources.
\end{abstract}

\maketitle


\section{\label{sec:intro}Introduction}

Thermodynamics establishes the fundamental laws governing state transformations of matter. However, when the system size approaches the microscopic scale, stricter conditions than those dictated by traditional thermodynamics must be satisfied \cite{Chubb2018beyondthermodynamic}. Within the framework of the resource-theoretic approach \cite{RevModPhys.91.025001rt}, significant progresses have been made over the past two decades \cite{Lostaglio_2019}. Generalized second laws have been formulated to describe both population dynamics \cite{Horodecki2013thermalmajor} and coherence dynamics \cite{arXiv:1708.04302} in quantum state evolution under thermal operations (TO)—a set of operations modeling interactions between the system and a thermal reservoir at a fixed temperature. Furthermore, necessary conditions for state transitions assisted by ancillary systems, such as catalysts \cite{RevModPhys.96.025005catalysis} and batteries \cite{second_law_battery_2024}, have also been established. The third law of thermodynamics has similarly been revisited and refined: not only have no-go theorems been rigorously proved through careful identification of relevant resources \cite{nc_third_law_2017,PRXQuantum.4.010332,PhysRevX.7.041033}, but the fundamental cooling limits achievable with finite resources have also been derived \cite{PhysRevLett.123.170605,PhysRevLett.134.070401,silva2024energy_cost_in_cooling,Henao2021catalytic}.

Heat-bath algorithmic cooling (HBAC) \cite{PNAS_HBAC} leverages quantum control techniques to transfer entropy from the target system to an auxiliary machine system, which subsequently dissipates the entropy into a thermal reservoir. This process enables cooling of the target system to temperatures below that of the surrounding heat bath. Beyond its foundational role in quantum thermodynamics, HBAC-based techniques have also been applied to improve sampling efficiency in quantum machine learning \cite{rodríguezbriones2025BQR}. Traditionally, the reset subsystem of the machine is assumed to undergo full thermalization with the heat bath \cite{ppa_schulman,PhysRevLett.94.120501_PPA,PhysRevLett.116.170501_PPA}. However, recent progresses demonstrate that improved cooling performance can be achieved by optimizing over partial thermalization protocols. These include thermalizing only specific subspaces, known as state reset (SR) \cite{Rodríguez-Briones_2017_SR}, and optimization over the full set of thermal operations, as in extended HBAC (xHBAC) \cite{Alhambra2019xHBAC}. Remarkably, xHBAC allows for asymptotic cooling of any finite-dimensional system to absolute zero, even in the absence of a machine. It is also observed that, because the reservoir is infinitely large, the complexity, measured by the dimension of the interaction unitary between the system and reservoir \cite{PRXQuantum.4.010332}, diverges.

In this paper, we address the following two questions:
\begin{itemize}
    \item Can cooling below the bath temperature be achieved without the use of a machine, under thermal operations of restricted complexity? 
    \item In the presence of a machine, can the achievable energy efficiency or cooling limit be improved?
\end{itemize}

For this purpose, we propose a set of thermal operations based on the collision model \cite{PhysRevLett.88.097905_collision_model,Strasberg2017,CICCARELLO20221}, where the heat bath consists of non-interacting identical molecules, and the system interacts sequentially with these molecules via energy-preserving unitaries. Crucially, we impose a constraint that each molecule is no larger than the system, ensuring that the overall operation remains of finite complexity. Within this framework, we analyze the condition on the system state before collision, construct a sub-bath cooling protocol based on the collision thermalization, and study the improvement of the energy efficiency and cooling limit by employing a qubit machine. Our main results are the following:

\begin{enumerate}
    \item For single collisions, we discuss when it is possible to cool the system below the bath's temperature (Sec.~\ref{ss:nogo}). We prove that this is impossible if there is an effective temperature, above the bath temperature or negative, defined for its initial state. This result follows from limiting the complexity of the thermal operations: when this complexity is unlimited, sub-bath cooling can be realized by the optimal $\beta$-permutation introduced in Ref.~\cite{Alhambra2019xHBAC}, which is an essential step in the xHBAC protocol. However, if the system state before collision lacks an effective temperature, sub-bath cooling may become possible, and we show an explicit case when it is.
    \item For the case where the system and the molecule have equally spaced and nondegenerate Hamiltonian, we construct an iterative protocol to cool the system below the bath temperature without the use of a machine (Sec.~\ref{subsec:wo_machine}).
    \item By introducing a single-qubit machine (Sec.~\ref{subsec:w_machine}), we show that the energy efficiency (quantified via the cumulative coefficient of performance, CoP), can be improved from zero to a strictly positive value. Moreover, if energy efficiency is not constrained, the inclusion of the machine also allows for an enhancement of the ultimate cooling limit.
\end{enumerate}

\begin{figure}
    \centering
    \includegraphics[width=0.8\linewidth]{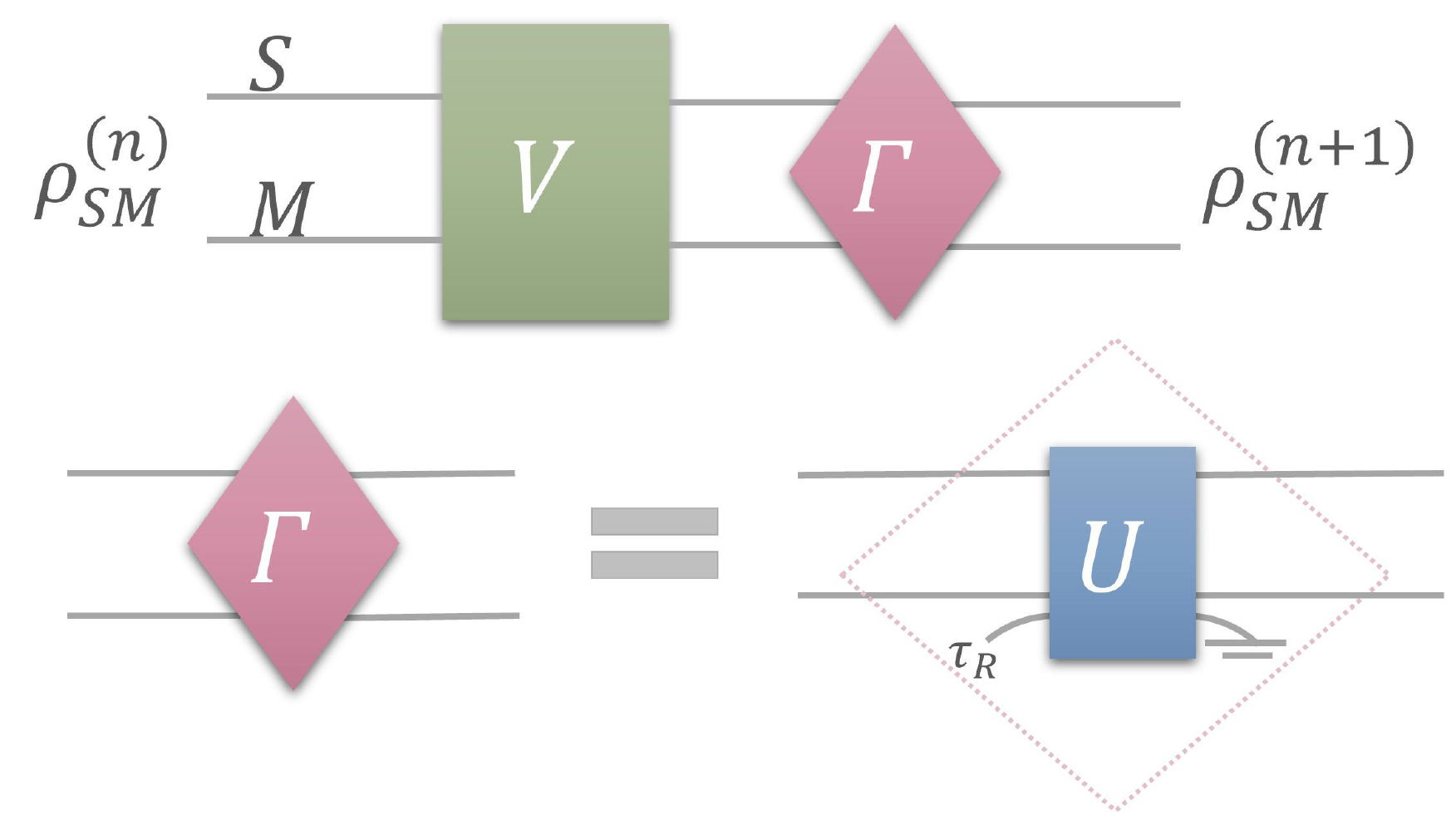}
    \caption{The configuration of one round in HBAC under coherent control. The target system, the machine and the heat bath are labeled as $S$, $M$, and $R$, respectively. The energy-non-preserving unitary $V$ is the recharging routine, and $\Gamma$, realized by interacting with a heat bath via a energy-preserving unitary $U$, is the thermalizing routine.}
    \label{fig:hbac}
\end{figure}

    \begin{table*}[ht]
    \centering
    \begin{tabular}{ |m{8cm}|>{\centering\arraybackslash}m{3cm}|>{\centering\arraybackslash}m{3cm}|>{\centering\arraybackslash}m{3cm}| }
    \hline
        Thermalization &  Machine & cooling limit & cumulative CoP\\
        \hline
        $\Gamma_1(\cdot)=\tr_M(\cdot)\otimes\tau(H_M,\beta)$ \cite{PhysRevLett.123.170605}  & $m$ qubits & $\beta^*=m\beta$ & positive\\
        \hline
        $\Gamma_2(\cdot)=\tr_m(\cdot)\otimes\tau(h,\beta)$ (PPA, \cite{PhysRevLett.116.170501_PPA}) &  $m$ qubits & $\beta^*=2^{m-1}\beta$ & positive \\
        \hline
        Full thermalization within a subspace of the composed system $SM$ (SR, \cite{Rodríguez-Briones_2017_SR}) &  $m$ qubits & $\beta^*=(2^{m+1}-1)\beta$ & zero\\
        \hline
        Full set of TO (xHBAC, \cite{Alhambra2019xHBAC}) &  w/o & $\beta^*\rightarrow\infty$ & zero\\
        \hline
        $\mathcal{C}_\beta(H^{(d_S)},H^{(d_r)},1)$ with $d_r\leq d_S$ (this work, Sec.~\ref{subsec:wo_machine}) &  w/o & $\beta^*\approx (d_r-1)\beta$ & zero\\
        \hline
        $\mathcal{C}_\beta(H^{(3)}+H^{(2)},H^{(3)},1)$ (this work, Sec.~\ref{subsec:w_machine})   & 1 qubit & $\beta^*=2\beta$ & positive\\
        \cline{3-4}
         &  & $\beta^*=4\beta$ & zero\\
         \hline
    \end{tabular}
    \caption{Performance of protocols with different thermalizing process, in the limit $N\rightarrow\infty$ of infinite number of steps. The cumulative coefficient of performance [CoP, Eq.~\eqref{eq:CoP}] is a measure of energy efficiency. It is zero, respectively positive, if the processes requires an unbounded, resp.~bounded, amount of energy.}
    \label{tab:comparison}
\end{table*}

\section{\label{sec:setting}thermal operations based on the collision model}
\subsection{definition}

In the collision model, the reservoir $R$ consists of a collection of identical non-interacting $d_r$-dimensional molecules, each with Hamiltonian $H_r$ and initially in the Gibbs state. The target system $S$ under consideration is a $d_S$-dimensional system with Hamiltonian $H_S$. The target system successively collides with the molecules, which is described by a energy-preserving unitary $U$ acting on the system and one molecule. Here and in the following, we use $\tau(H,\beta):=e^{-\beta H}/\tr(e^{-\beta H})$ to denote the Gibbs state of a system with Hamiltonian $H$ and at inverse temperature $\beta$. Also, for simplicity, we label the Gibbs state of the molecule and the system as $\tau^r\equiv\tau(H_r,\beta)$ and $\tau^S\equiv\tau(H_S,\beta)$, respectively, and the corresponding energy distributions as $\{\tau_j^r\}_{j=0}^{d_r-1}$ and $\{\tau^S_k\}_{k=0}^{d_S-1}$.

Based on the collision model, we propose a set of thermal operations defined as
\begin{eqnarray}
    \mC_\beta(H_S,H_r,N)=\left\{\Lambda^{\circ N}|\Lambda(\cdot)=\tr_r[U(\cdot\otimes\tau^r)U^\dagger],\right.\nonumber\\
    \left.[U,H_S+H_r]=0\right\}.
\end{eqnarray} Notice that $U$ is the running parameter defining the set, all the other objects being fixed. According to Ref.~\cite{PRXQuantum.4.010332}, the complexity of a quantum operation can be measured by the effective dimension of the unitary (defined as the dimension of the subspace upon which the unitary acts nontrivially, and upper bounded by the dimension of the unitary) in realizing it, and the time cost of an operation can be measured by the number of unitary operations. Hence in our case, the time cost of an operation in $\mC_\beta(H_S,H_r,N)$ is $N$, and the complexity of it is upper bounded by the dimension of $U$, which is $d_Sd_r$. When both the system and the molecule are finite-dimensional, the complexity in $\mC_\beta(H_S,H_r,N)$ is finite. In our model, the molecules in the bath are identical, and the unitary in each collision is also the same (a larger class of operations could be achieved by changing the interaction times or even the interaction in every collision, at the cost of having to control such parameters). Importantly, we further require that the dimension of each molecule is not larger than that of the target system.

\subsection{Comparison with Markov thermal processes}

When the time step $\Delta t$ of each collision is infinitesimal, the quantum processes described by the collision model are closely related to the Markov processes. On the one hand, any operation described by the collision model is divisible into single collisions, and is thus Markov. On the other hand, multipartite collision models introduced in Ref.~\cite{PhysRevLett.126.130403mcm} can reproduce any Gorini-Kossakowski-Sudarshan-Lindblad master equation, given that the bath consists of qubits with different energy gaps and that the interaction Hamiltonian in each collision is carefully designed.

However, in the definition of $\mC_\beta(H_S,H_r,N)$, the energy-preserving unitary $U$ that describes the interaction in each collision is not necessarily realizable within an infinitesimal time interval, so the relation between the thermal operations based on collision model $\mC_\beta(H_S,H_r,N)$ and the set of thermal operations which are solutions to Markov master equations (Markov thermal operations, $\MTO_\beta$ \cite{PhysRevA.106.012426,PhysRevLett.129.040602}) is not straightforward. We discuss the state transitions allowed in the two models using the concept of the free operation cone. Let $\mathcal{FO}$ be a set of operations and $\rho$ be a given input state: The $\mathcal{FO}$-cone of $\rho$ is defined as the set of output states which can be obtained by operations in $\mathcal{FO}$
\begin{equation}
    \cone_{\mathcal{FO}}(\rho)=\{\Lambda(\rho)|\Lambda\in\mathcal{FO}\}.
\end{equation}
This cone is not a convex set, because having bounded the ancilla's dimension, extra classical randomness is not free; and so one cannot realize the mixture $p_1\Lambda_1(\rho)+p_2\Lambda_2(\rho)$ by tossing a biased coin and implementing either $\Lambda_j$.

We study two cases, the second proving (Fig.~\ref{fig:cone}) that the two sets of operations are in general incompatible, i.e.~some states can be reached by one and not the other.

\subsubsection{Both the system and the molecule are a single qubit}

Both the system and the molecule are a single qubit, with the same Hamiltonian $H_S=H_r=H^{(2)}\equiv E|1\rangle\langle1|$. 

\begin{theorem}\label{th:cone}
For any qubit state $\rho$ and number of collisions $N$,
    \begin{equation}
        \cone_{\mC_\beta(H^{(2)},H^{(2)},N)}(\rho)\subseteq\cone_{\mC_\beta(H^{(2)},H^{(2)},N+1)}(\rho),\label{eq:cone_n}
    \end{equation}
    and
        \begin{equation}
        \cone_{\mC_\beta(H^{(2)},H^{(2)},\infty)}(\rho)=\cone_{\MTO_\beta}(\rho).\label{eq:cone_MTO_col}
    \end{equation}
    Besides, the convex hull of $\cone_{\mC_\beta(H^{(2)},H^{(2)},1)}(\rho)$ equals to $\cone_{\MTO_\beta}(\rho)$.
\end{theorem}

The proof is in Appendix~\ref{app:qubit_cone}. Several discussions are in order.

Firstly, although more states can be achieved with increasing number of collisions as shown in Eq.~(\ref{eq:cone_n}), any extreme point of $\cone_{\mC_\beta(H^{(2)},H^{(2)},\infty)}(\rho)$ can be achieved by a single collision. 
It indicates that, if one aims at simulating the qubit state transition from a given state $\rho$ to any state at the extreme point of $\cone_{\MTO_\beta}(\rho)$, instead of reproducing the time evolution equation, the time cost, measured by $N$, can be reduced to 1.

Besides, Theorem~\ref{th:cone} implies that the state transformation under $\mC_\beta(H^{(2)},H^{(2)},N)$ is realizable by a Markov thermal process, no matter how strong the interaction is in each collision. Although in general, the effective time evolution of the system state when colliding with a molecule cannot be reproduced by a Markov equation due to the strong interaction, the input-output relation induced by $\mC_\beta(H^{(2)},H^{(2)},N)$ can be realized by MTO. In other words, the memory in a single-qubit molecule is not large enough to induce state transitions which cannot be realized by Markov thermal processes.

Further, if the system is $d_S$-dimensional and has non-degenerate energy gaps, and the bath consists of qubit molecules with different energy gaps, which can resonant with different energy gaps in the system, Theorem \ref{th:cone} indicates that the state obtained by colliding with the molecules cannot exceed the MTO cone, no matter how strong the interaction is or how large the energy gap is in the molecules.

\subsubsection{Beyond two qubits}

In general, $\mC_\beta(H,H,N)$ and $\MTO_\beta$ are incompatible. This can already be shown by looking at population dynamics. As an example, let us consider the case where both the system and the molecule are qutrits with equally spaced Hamiltonians: $H_S=H_r=H^{(3)}\equiv\sum_{j=0}^{2}jE|j\rangle\langle j|$. The population distribution of the Gibbs state $\tau(H^{(3)},\beta)$ is $\vec{\tau}=(\tau_0,\tau_1,\tau_2)$ with $\tau_j=q^j/(1+q+q^2)$ and $q\equiv e^{-\beta E}$.
For an initial system state with population distribution $\vec{p}=(\tau_1,\tau_0,\tau_2)$, the $\mC_\beta(H^{(3)},H^{(3)},1)$-cone as well as the $\MTO_\beta$-cone are depicted in Fig.~\ref{fig:cone}. (Detailed calculation of the $\mC_\beta(H^{(3)},H^{(3)},1)$-cone is in Appendix \ref{app:qutrit_cone}.) Clearly, the two cones are incompatible with each other: while the minimum value of $p'_2$ in the output via $\MTO_\beta$ is lower than that via $\mC_\beta(H^{(3)},H^{(3)},1)$, $\mC_\beta(H^{(3)},H^{(3)},1)$ can achieve higher maximum values of $p_0'$ and $p'_2$ than $\MTO_\beta$. This incompatibility indicates that, qutrit molecules are large enough to exceed the Markov constraint. 

\begin{figure}
    \centering
    \includegraphics[width=0.8\linewidth]{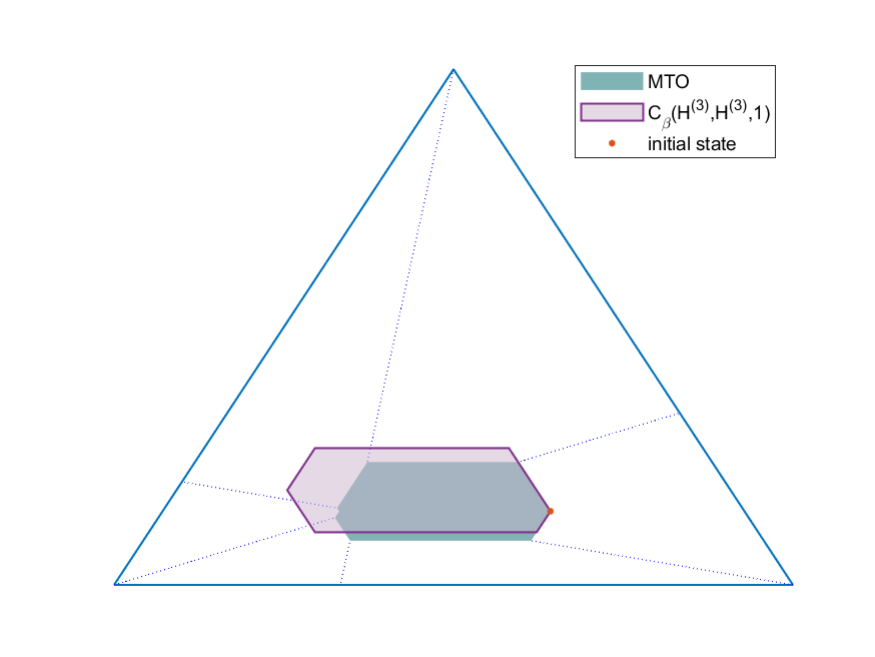}
    \caption{The set of reachable states from initial state $\vec{p}=(\tau_1,\tau_0,\tau_2)$. Here we choose $q=0.5$.}
    \label{fig:cone}
\end{figure}

\section{Single-collision cooling}
\label{ss:nogo}

In this Section, we consider single-collision cooling, focusing in particular on the possibility of cooling the system below the temperature of the bath. We first show that this is impossible if the state of the system can be associated to an effective temperature $\beta'$ (we shall study $-\infty\leq \beta'\leq \beta$: if $\beta'>\beta$, the system is already cooler than the bath and the question makes little sense). Then we show that relaxing this condition can lead to cooling below the bath's temperature.

\subsection{No-go theorems}

Intuitively, if the target system is at a higher temperature than the environment, and the molecule is no larger than the target system, one would not expect to cool the system below the environment temperature by colliding with a molecule. Here we formulate this argument and strictly prove it.

Let $H_S=\sum_{k=0}^{d_S-1}E_k^S|k\rangle\langle k|$ with $E^S_k\leq E^S_{k'},\forall k\leq k'$ be the Hamiltonian of the target system and $H_r=\sum_{j=0}^{d_r-1}E_j^r|j\rangle\langle j|$ with $E^r_j\leq E^r_{j'},\forall j\leq j'$ be the Hamiltonian of a molecule.

Before the collision, the target system is in state $\rho$ with energy distribution $\vec{p}=(p_0,\dots,p_{d_S-1})^\mathrm{T}$. While $\rho$ may have coherences and even be pure, to study cooling we need only this distribution, because for phase-covariant channels, like the ones under study, the evolution of the populations is independent of the coherences \cite{PhysRevX.5.021001,PhysRevLett.115.210403}. The molecule is initially in the Gibbs state $\tau^r=\tau(H_r,\beta)$. 
During the collision, a joint unitary $U$ satisfying the energy preservation condition $[U,H_S+H_r]=0$ is applied to the composed system of $S$ and $r$. Notice that $H_S$ and $H_r$ must share some energy gaps for such a $U$ to couple system and reservoir; in the absence of common gaps, our setting leads to trivial dynamics.

The aim of cooling is to maximize the ground state population of $S$ in the output. Hence, the figure-of-merit in one collision is
\begin{equation}
    p_0^{(1)*}=\max_{U:[U,H_S+H_r]=0}\langle 0_S|\tr_r[U(\rho\otimes\tau^r)U^\dagger]|0_S\rangle.
\end{equation}
Let $\tau^S_0=\langle0_S|\tau(H_S,\beta)|0_S\rangle$ be the ground state population for $S$ at temperature $\beta$. We say that the system cannot be cooled below the environment temperature if $p_0^{(1)*}\leq\tau^S_0$.

We first consider the following three requirements:
\begin{itemize}
    \item[(R1)] $d_S\geq d_r$. This is the condition that the molecule is no larger than the target system.
    \item[(R2)] $E_j^S=E_j^r$, $\forall j=0,\dots,d_r-1$. This says that the spectra of $H_S$ and $H_r$ are equal for the lowest $d_r$ energy levels, ensuring that the gaps are also equal.
    \item[(R3)] $p_k e^{\beta E_k^S}\leq p_{k'}e^{\beta E_{k'}^S}$, $\forall 0\leq k\leq k'\leq d_S-1$. When this condition is satisfied, the state of the system can be associated with an effective temperature $-\infty\leq\beta'\leq\beta$ which is no lower than the environment temperature, or negative (see Appendix \ref{app:beta_ordering}); in particular, Gibbs state $\tau(H_S,\beta')$ satisfy it.
\end{itemize}


Now we are ready to present our first no-go theorem. The proof is in Appendix \ref{app:no_go}.

\begin{theorem}\label{th:no_go}
    If (R1)-(R3) are satisfied, $p^{(1)*}_0\leq\tau^S_0$.
\end{theorem}

Also, the preliminary that ``the molecular is no larger than the system'' can also be interpreted as follows. The target system consists of $\mu$ copies of independent particles each with Hamiltonian $h=\sum_l\epsilon_l|l\rangle\langle l|$, while the molecule consists of $\nu(\leq \mu)$ copies of the same particle.  That is, $H_S=\sum_{\eta=1}^{\mu}H_\eta^\mu$ and $H_r=\sum_{\eta=1}^{\nu}H_\eta^\nu$, where $H_\eta^\mu=\underbrace{\iden\otimes\dots\otimes\iden}_{\eta-1}\otimes h\otimes\underbrace{\iden\otimes\dots\otimes\iden}_{\mu-\eta}$ and $H_\eta^\nu$ is defined similarly. Without loss of generality, we set $\epsilon_0=0$. For this case, (R1) naturally holds, but (R2) no longer holds if $\mu>\nu$. However, we are still able to prove the no-go theorem, as long as the initial state satisfies (R3). Refer to Appendix \ref{app:no_go1} for the proof.

\begin{theorem}\label{th:no_go1}
    If the target system $S$ consists of $\mu$ copies of independent particles, the molecule consists of $\nu(\leq \mu)$ copies of the same kind of particles, and the initial state of $S$ satisfies (R3), then $p_0^{(1)*}\leq\tau_0^S$.
\end{theorem}

It is worth noticing that, (R3) includes any population-inversion state with $p_0\leq p_1\leq\dots\leq p_{m-1}$ as a special case. Such states are called the most activated states in the xHBAC protocol, see Ref. \cite{Alhambra2019xHBAC} for more details. In each round of xHBAC, the target system is prepared to a population-inversion state by a unitary operation, before it interacts with the bath. A direct consequence of our no-go theorems is that, the xHBAC protocol cannot be generalized directly to the situation where the part of the bath which interacts with target system is no larger than the size of the system.

\subsection{Possibility of cooling below the bath's temperature}

If the state of the system initially \textit{does not} possess an effective temperature, i.e.~if it is ``out-of-equilibrium'' in the sense that (R3) does not hold, then the system can be cooled below the environment temperature by colliding with a molecule of the same size as $S$. 

An example is as follows. The system $S$ consists of three qubits, labeled as $A_1,A_2$ and $A_3$; the molecule also consists of three qubits, labeled as $B_1,B_2$ and $B_3$; all the qubits have the same Hamiltonian $h=\epsilon |1\rangle\langle 1|$. Initially, qubit $A_1$ is in the state $\rho=\bar{p}_0|0\rangle\langle 0|+(1-\bar{p}_0)|1\rangle\langle 1|$, while the initial states of the other five qubits are the Gibbs state $\bar{\tau}=\tau(h,\beta)$, where $\bar{p}_0<\bar{\tau}_0=\langle 0|\bar{\tau}|0\rangle$. Thus, the ground state of the system has population $\bar{p}_0\bar{\tau}_0^2<\bar{\tau}_0^3\equiv \tau^S_0$, the latter being the ground state population of $\tau(H_S,\beta)$. Moreover, for eigenstates $|100\rangle$ and $|011\rangle$, the initial populations are respectively $p_{100}=\bar{p}_1\bar{\tau}_0^2$ and $p_{011}=\bar{p}_0\bar{\tau}_1^2$, while the energy are respectively $E_{100}=\epsilon$ and $E_{011}=2\epsilon$. Then (R3) is violated because $p_{100}e^{\beta E_{100}}=p_1\bar{\tau}^2e^{\beta\epsilon}>p_0\bar{\tau}^2=p_{011}e^{\beta E_{011}}$ but $E_{100}<E_{011}$. 

Under these conditions, the ground state population of the system after one collision can be made larger than $\tau^S_0$, with a suitable choice of energy-preserving $U$. An intuitive protocol is to cool $A_1$ by interacting it with $r=B_1B_2B_3$ while keeping the state of $A_2A_3$ untouched. The ground state population for $A_1$ becomes $\bar{p}_0^{(1)}=\bar{\tau}_0+\bar{\tau}_0\bar{\tau}_1(\bar{\tau}_0-p_0)>\bar{\tau}_0$, and therefore the ground state population of $S=A_1A_2A_3$ reaches $p^{(1)}=\bar{p}_0^{(1)}\bar{\tau}_0^2>\tau^S_0$.


Another example can be read out of Fig.~\ref{fig:cone}. There, both the system and the molecule are one qutrit with equally spaced Hamiltonians $H_S=H_r=H^{(3)}\equiv\sum_{j=0}^2 jE|j\rangle\langle j|$. For initial state with population distribution $\vec{p}=(\tau_1,\tau_0,\tau_2)$, the ground state population in the output can reach $p_0^{(1)*}=\tau_0+\tau_1(\tau_0-\tau_1)>\tau_0$ (see Appendix \ref{app:beta_ordering} for the calculation).

This example motivates us to construct an iterative protocol for cooling based on finite-complexity thermalization without the use of a machine.

\section{Iterative cooling protocols}

In this Section, we first briefly review HBAC protocols under coherent control, and see how the control over thermalization improves the cooling limit. Then we present two iterative protocols based on collisions with molecules not larger than the system, one without machine, the other using a qubit machine.

\subsection{A unified view of Heat-bath Algorithmic Cooling (HBAC)}

In a HBAC protocol, a machine $M$ works cyclically between the target system $S$ and the heat bath $R$, in order to refrigerate the target system, usually below the temperature of the heat bath. If the energy necessary for the cooling process comes from a unitary which pumps the systems (the target, the machine, or the composition of both) to a state with higher energy, the cooling protocol belongs to the coherent control scenario.

In general, a coherent control HBAC process consists of $N$ rounds, each made of two routines: the recharging routine and the thermalizing routine (Fig.~\ref{fig:hbac}). In the recharging routine, a unitary $V$ is applied to the target system $S$ and the machine $M$, providing the energy required for cooling. In the thermalizing routine, $S$ and $M$ are brought to interact with a bath $R$ via an energy-preserving unitary $U$, which satisfies $[U,H_S+H_M+H_R]=0$, such that the entropy initially contained in $S$ is transferred to the bath.

The performance of a cooling protocol is usually characterized by ground state population in the output. If the asymptotic state to which the target system converges is in the form of a Gibbs state $\rho^*=\tau(H_S,\beta^*)$, then the cooling performance can also be characterized by the effective inverse temperature $\beta^*$.

The cooling limit is affected by the ability to control the thermalization routine. For example, let the target system be a qubit with Hamiltonian $h$, and the machine consist of $m$ independent identical qubits, each with Hamiltonian $h$. If $\Gamma=\Gamma_1$ is full thermalization of the machine $\Gamma_1(\cdot)=\tr_M(\cdot)\otimes\tau(H_M,\beta)$ \cite{PhysRevLett.123.170605}, then the reachable cooling limit reads $\beta^*_1=m\beta$. If $\Gamma=\Gamma_2$ is full thermalization of the $m$th qubit in the machine $\Gamma_2(\cdot)=\tr_m(\cdot)\otimes\tau(h,\beta)$ \cite{PhysRevLett.116.170501_PPA}, then the reachable cooling limit reads $\beta^*_2=2^{m-1}\beta$. If $\Gamma=\Gamma_3$ is the so-called state reset, which acts on the composed system of $S$ and $M$ thermalizing the populations only in given subspace while preserving the populations on other energy levels \cite{Rodríguez-Briones_2017_SR}, the cooling limit $\beta^*_3=(2^{m+1}-1)\beta$ can be reached. If the thermalization $\Gamma_4$ is optimized over the full set of thermal operations (TO), as in xHBAC \cite{Alhambra2019xHBAC}, then absolute zero $\beta_4^*\rightarrow\infty$ can be reached even without a machine. Therefore, as the control over the thermalizing routine is enhanced, the reachable cooling limit is improved.

The heat bath in xHBAC is assumed to be infinite. Actually, finite-sized heat bath can set limitations on state transitions \cite{PhysRevE.97.062132,nc_finite_bath_work}, and hence, on the efficiencies of tasks such as cooling or heat engine \cite{Pozas-Kerstjens_2018}. In particular, when the size of the heat bath is finite, the $\beta$-swap operation which is employed in xHBAC is not realizable \cite{Scharlau2018quantumhornslemma}. It is shown that \cite{Scharlau2018quantumhornslemma}, for any input state of a qubit system, the ground state population cannot reach one by interacting with a finite-size bath.

\subsection{Protocols in the absence of a machine \label{subsec:wo_machine}}

Now we construct an iterative protocol of cooling for the following setting. The Hamiltonian of $S$ and $r$ are both equally spaced and non-degenerate, i.e., $H_S=H^{(d_S)}$ and $H_r=H^{(d_r)}$, where $H^{(d)}\equiv \sum_{j=0}^{d-1}j E|j\rangle\langle j|$. For latter convenience, we label $q\equiv e^{-\beta E}$. Again, we are interested in cooling $S$ with molecules no larger than $S$, so $d_S\geq d_r$.

The protocol is depicted in Fig. \ref{fig:hbac} in the absence of the machine, i.e., both the unitary $V$ and the partial thermalization $\Gamma$ act on the target system $S$. Here $\Gamma$ is realized by a single collision with a molecule, $\Gamma\in\mC_\beta(H^{(d_S)},H^{(d_r)},1)$. Importantly, the molecule here is by no means equivalent to the machine in the protocol where a machine is brought to interact with the system via a energy-non-preserving unitary and then fully thermalized in the environment: rather, as a part of the environment, the molecule interacts with the target system via energy-preserving unitary operations. In addition, $V$ and $\mE$ are fixed for each round, instead of being optimized for each round. This requirement has also been employed in Ref. \cite{PhysRevLett.122.220501} to make the protocol more implementable.

By constructive proof, we show that the ground state population in the output converges to
\begin{equation}
    p_0^*=\left\{\begin{array}{cc}
        \tau_0\left(H_S,(d_r-1)\beta\right), & d_r=3, \\
        \tau_0\left(H_S,(d_r-1)\beta\right)-o(q^{d_r+2}), & d_r\geq4,
    \end{array}\right.
\end{equation}
where $\tau_0\left(H_S,(d_r-1)\beta\right)=\langle 0|\tau\left(H_S,(d_r-1)\beta\right)|0\rangle$ is the ground state population of the Gibbs state at inverse temperature $(d_r-1)\beta$.
It means that, in the asymptotical limit, the inverse temperature $(d_r-1)\beta$ can be approached for $d=3$, and approximately approached for $d\geq4$ when $q$ is small.

We present the protocol which achieves the above cooling limit for $d_S=d_r=3$; the protocols for higher $d_S$ and $d_r$ are similar, and can be found in Appendix \ref{app:protocol}. The total Hamiltonian of $S$ and $r$ is
\begin{equation}
    H_{Sr}=E\Pi_1+2E\Pi_2+3E\Pi_3+4E|22\rangle\langle 22|,
\end{equation}
where $\Pi_1=|01\rangle\langle 01|+|10\rangle\langle10|$, $\Pi_2=|02\rangle\langle 02|+|11\rangle\langle11|+|20\rangle\langle20|$, and $\Pi_3=|12\rangle\langle12|+|21\rangle\langle21|$ are projections to the energy eigenspaces associated with $E$, $2E$, and $3E$, respectively. In the recharging routine of each round, the target goes through a unitary operation $\mathcal{V}(\cdot)=V(\cdot)V^\dagger$ with $V=|0\rangle\langle 1|+|1\rangle\langle0|+|2\rangle\langle2|$. The thermalizing routine is realized by an energy-preserving collision $U$ with a molecule $r$ initially in the Gibbs state $\tau^r$, inducing the transition matrices $G_j$ in the energy eigenspace of $H_{Sr}$ associated with $jE$. Explicitly,
\begin{equation}
    G_1=\left(\begin{array}{cc}
        0 & 1 \\
        1 & 0
    \end{array}\right), G_2=\left(\begin{array}{ccc}
        0 & 1 & 0 \\
        0 & 0 &1 \\
        1 & 0 & 0
    \end{array}\right),
    G_3=\iden_2.
\end{equation}
The effective operation of the thermalizing routine on $S$ is then $\Gamma(\cdot)=\tr_r[U(\cdot\otimes\tau^r)U^\dagger]$. After some calculations, we obtain the transition matrix on $S$ of each round
\begin{equation}
    G_{\mE\circ\mathcal{V}}=\left(\begin{array}{ccc}
        \tau^r_0+\tau^r_1 & \tau^r_0 & 0 \\
        \tau^r_2 & \tau^r_1 &  \tau^r_0 \\
        0 & \tau^r_2 &  \tau^r_1+\tau^r_2
    \end{array}\right).\label{eq:G_3}
\end{equation}
The unique fixed point for $G_{\mE\circ\mathcal{V}}$ as in Eq. (\ref{eq:G_3}) is
\begin{equation}
    \vec{p}_*=\frac{1}{1+q^2+q^4}(1,q^2,q^4)^\mathrm{T}.
\end{equation}
Hence after $N\rightarrow\infty$ rounds, the energy distribution of the target can approach $\vec{p}_*$ for any input state.

Precisely, we calculate the distribution after the $N$th round as
\begin{equation}
    \vec{p}^{(N)}=\vec{p}_*+(2\tau^r_1)^{N-1}\vec{\delta},
\end{equation}
where $\vec{\delta}=[q(p_0^{(0)}-p_0^*)-(p_2^{(0)}-p_2^*)](q,-1+q,-q)^{\mathrm{T}}$.
Therefore, $\vec{p}^{(N)}$ converges exponentially to $\vec{p}_*$ at the rate $2\tau^r_1=\frac{2q}{1+q+q^2}$.

\subsection{Protocols with a qubit machine}
\label{subsec:w_machine}

Finally we consider an iterative protocol with the smallest possible machine, a qubit with Hamiltonian $H_M=H^{(2)}$. The system and the molecule are both qutrits with Hamiltonian $H^{(3)}$. We compare this protocol (Protocol II) with the one without machine (Protocol I) just presented in Sec. \ref{subsec:wo_machine}. We show that the cooling protocol with the smallest machine outperforms those without a machine in both energy efficiency and cooling limit in the asymptotical scenario. In the main text we focus on the iterative protocols; for single-round protocols, the qubit machine cannot improve the achievable ground state population but can reduce the consumed energy (Appendix \ref{app:w_machine}).

\subsubsection{Improved energy efficiency}
In order to characterize the energy efficiency of cooling protocols, we employ the classical coefficient of performance (CoP) \cite{PhysRevA.110.022215}, which is defined as
\begin{equation}\label{eq:CoP}
    K=\frac{-\Delta U}{W},
\end{equation}
where $W$ is the consumed energy and $-\Delta U$ is the amount of energy taken away from the system. For iterative protocols, we are interested in the cumulative CoP. Let $\vec{p}_X^{(n)}$ be the vector of populations of system $X$ after the $n$-th round, and $\vec{H}_X$ denote the vector of energy eigenvalues of system $X$. Then $W^{(n)}=(V\vec{p}_{SM}^{(n-1)}-\vec{p}_{SM}^{(n-1)})\cdot\vec{H}_{SM}$ is the work consumption in the $n$-th round, and $-\Delta U^{(n)}=(\vec{p}^{(n-1)}_S-\vec{p}_S^{(n)})\cdot\vec{H}_S$ is the amount of energy reduction of the system $S$ in the $n$-th round. The cumulative CoP up to $N$ rounds is
\begin{equation}
    K^{(N)}=\frac{-\sum_{n=1}^N\Delta U^{(n)}}{\sum_{n=1}^N W^{(n)}}.
\end{equation}

\begin{figure}
    \centering
    \includegraphics[width=0.7\linewidth]{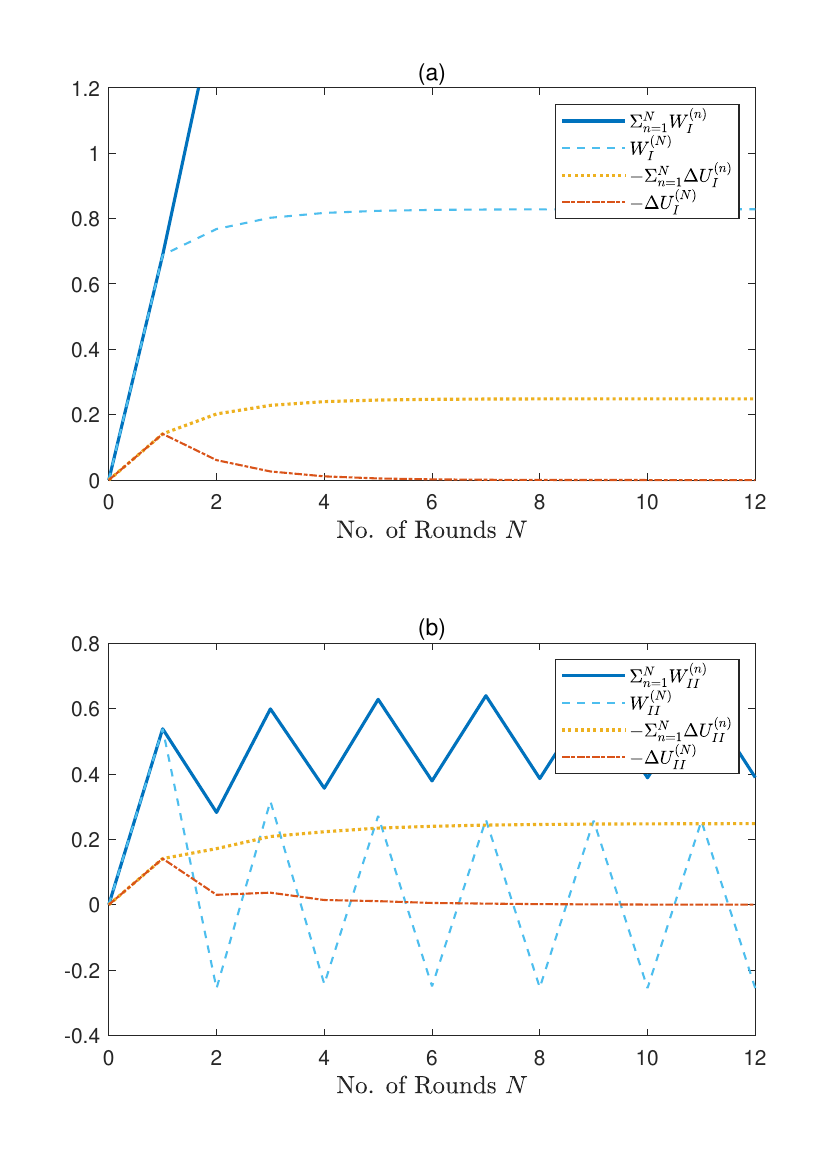}
    \caption{The work consumption and the energy reduction for both Protocol I (Panel (a)) and Protocol II (Panel (b)). Here we choose $q=0.3$ and $E=1$.}
    \label{fig:work_con}
\end{figure}

For Protocol I, $K^{(N)}$ declines to zero as $N$ grows for the following reason. Without a machine, the unitary $V$ acts solely on the system $S$. In order to cool the system to a lower temperature, the system has to be driven to a state in a different $\beta$-ordering, which means that the energy consumption $W^{(n)}$ in each step is lower bounded by some non-vanishing positive value $w$.
Meanwhile, the total energy taken away from the system is upper bounded by its initial mean energy $U_0$. See Fig. \ref{fig:work_con} (a) for an example. It follows that
\begin{equation}
    K^{(N)}_{\mathrm{I}}\leq\frac{U_0}{Nw},
\end{equation}
which approaches zero for large $N$.
The vanishing CoP for large $N$ is also observed for xHBAC in Ref. \cite{PhysRevA.110.022215}, and the reason is similar.

In order to overcome this, we employ a qubit machine, and construct the following protocol. In the recharging routine, the bit-flip unitary $\sigma^x$ acts locally on the machine, i.e., $V=\iden_S\otimes\sigma^x_M$. In the thermalizing routine, the energy preserving unitary $U$, which acts collectively on the composed system of $SMr$, permutes the populations on $|002\rangle$ and $|110\rangle$, as well as those on $|102\rangle$ and $|210\rangle$, and keeps the other populations unchanged. See Appendix \ref{app:protocol} for the analytic expression of the transition matrix and some discussions on it.

\begin{figure}
    \centering
    \includegraphics[width=0.8\linewidth]{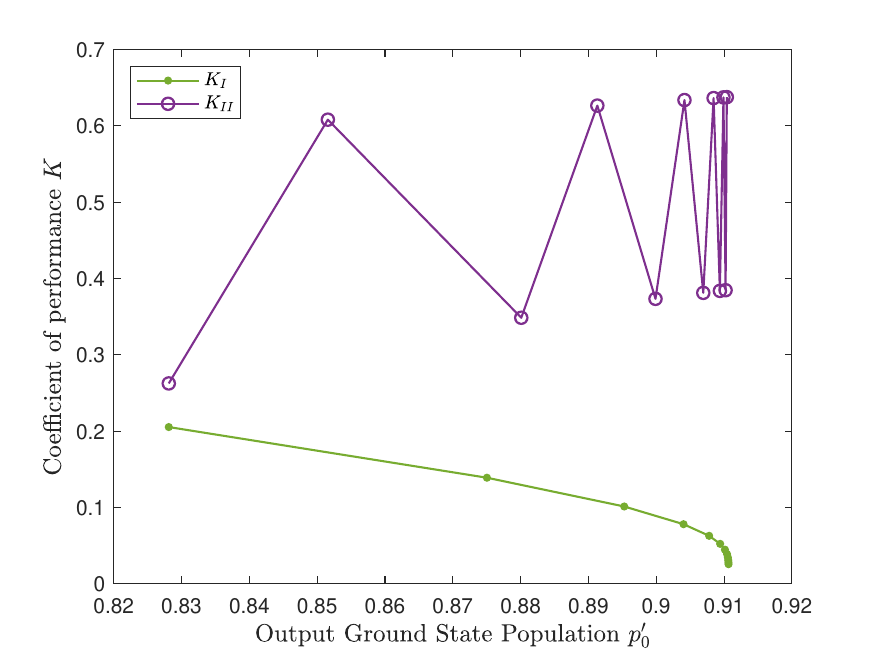}
    \caption{Comparison between the cumulative CoP of Protocols I and II. Here we choose $q=0.3$.}
    \label{fig:cop}
\end{figure}

With this, the work consumption $W^{(N)}_{\mathrm{II}}$ and energy reduction $-\Delta U^{(N)}_{\mathrm{II}}$ in the $N$-th round, as well as the cumulative work consumption and energy reduction, are depicted in Fig. \ref{fig:work_con} (b). The oscillation behavior of the consumed work in each round is observed: while for $(2k+1)$-th round, certain amount of work is consumed, for $(2k)$-th round, $W^{(2k)}_{\mathrm{II}}<0$, meaning that not only no work is consumed, but also certain amount of work is extracted. As a result, the cumulative work consumed in Protocol II is bounded from above by some finite value. This leads to a positive lower bound on cumulative CoP, as shown in Fig. \ref{fig:cop}.

\subsubsection{Improved cooling limit\label{subsub:imp_cooling_limit}}
The cooling limit can also be improved by employing a machine, if we put no restrictions on the energy efficiency. 
The protocol is as follows.

In the recharging routine, a Pauli-$x$ unitary acts on the machine, i.e., $\tilde{V}=\iden_S\otimes\sigma^x_M$. The thermalization routine $\tilde{\Gamma}$ is detailed in Appendix \ref{app:w_machine}. The transition matrix associated with the operation $\tilde{\Gamma}\circ\tilde{V}$ on $SM$ in one round reads
\begin{equation}
    G_{\tilde{\Gamma}\circ\tilde{V}}=\left(\begin{array}{cccccc}
        \tau^r_0+\tau^r_1 & \tau^r_0 &  &  &  & \\
        \tau^r_2 & \tau^r_1 & \tau^r_0 & & & \\
         & \tau^r_2 & \tau^r_1 & \tau^r_0 & & \\
         & & \tau^r_2 & \tau^r_1 & \tau^r_0 & \\
         & & & \tau^r_2 & \tau^r_1 &\tau^r_0 \\
         & & & & \tau^r_2 & \tau^r_1+\tau^r_2
    \end{array}\right),
\end{equation}
the fixed point of which is 
\begin{equation}
    \frac{1}{\mathcal{N}}\left(\begin{array}{c}
        1 \\
        q^4 \\
        q^8
    \end{array}\right)_S\otimes\left(\begin{array}{c}
        1 \\
        q^2
    \end{array}\right)_M,
\end{equation}
where $\mathcal{N}$ is the normalization factor. It means that, the state of the system converges to a state with inverse temperature $4\beta$, in contrast to the scenario without a machine, where the cooling limit is $2\beta$.

\section{Conclusion}
We have proposed a set of thermal operations of finite complexity based on the collision model, and applied these operations in the thermalizing routine of algorithmic cooling. If an effective temperature can be defined for the initial system state and it is no lower than the bath temperature or negative, we prove that the system cannot be cooled below the environment temperature. 

However, the cooling opportunity opens up for initial states without a temperature. Specifically, if the system is initially in a non-equilibrium state such that is not in the same $\beta$-ordering as a Gibbs state, we show that the ground state population of the system can be made larger than that of the Gibbs state at the environment temperature, by a single-collision with a qutrit molecule. Based on this observation, we construct an iterative protocol to cool the system below the bath temperature, for the case where the system and the molecule have equally spaced and nondegenerate Hamiltonian. Specifically, for single-qutrit molecules, the system's inverse temperature can asymptotically approach $2\beta$. More generally, for $d_r$-dimensional molecules, the inverse temperature $(d_r-1)\beta$ can be approximately attained in the asymptotic limit. 

In addition, we show that by introducing a single-qubit machine, the energy efficiency—quantified via the cumulative coefficient of performance (CoP)—can be improved from zero to a strictly positive value. Moreover, if energy efficiency is not constrained, the inclusion of the machine also allows for an enhancement of the ultimate cooling limit.

\begin{acknowledgments}

We thank Jeongrak Son and  Nelly H.Y. Ng for helpful discussions. This project is supported by the National Research Foundation, Singapore through the National Quantum Office, hosted in A*STAR, under its Centre for Quantum Technologies Funding Initiative (S24Q2d0009); by the Ministry of Education, Singapore, under the Tier 2 grant ``Bayesian approach to irreversibility'' (Grant No.~MOE-T2EP50123-0002); by NSFC under Grant No. 11774205, and the China Scholarship Council (award to Xueyuan Hu for 1 year’s study abroad at the National University of Singapore).
\end{acknowledgments}

\appendix

\section{Proof of Theorem \ref{th:cone} \label{app:qubit_cone}}
The system qubit initially in the state with the Bloch vector $(\eta\cos\varphi,\eta\sin\varphi,z)$ can be transformed to a state $(\eta'\cos\varphi',\eta'\sin\varphi',z')$ via $\mC_\beta(H^{(2)},H^{(2)},N)$, if and only if 
\begin{equation}
    \eta z^N_\tau\sqrt{\frac{z'-z_\tau}{z-z_\tau}}\leq \eta' \leq\eta\sqrt{\frac{z'-z_\tau}{z-z_\tau}},\label{eq:cone_qubit}
\end{equation}
for $z\neq z_\tau$, and $\eta'\in[0,\eta]$ for $z=z_\tau$. Here $z_\tau=\frac{1-q}{1+q}$ is the $z$-component of the Gibbs state $\tau(H^{(2)},H^{(2)},\beta)$ and $q\equiv e^{-\beta E}$. 
This $\mC_\beta(H^{(2)},N)$-cone is derived from the result in Ref.~\cite{PhysRevLett.88.097905_collision_model} as follows.

Before the collision, the state of the system is
\begin{equation}
    \rho=\left(\begin{array}{cc}
        \rho_{00} & \rho_{01} \\
        \rho_{10} & \rho_{11}
    \end{array}\right),
\end{equation}
where $\rho_{00,11}=\frac{1\pm z}{2}$, and $\rho_{01}=\rho_{10}^*=\eta e^{-i\varphi}$. The molecule qubit is in the Gibbs state $\tau=\tau_0|0\rangle\langle0|+\tau_1|1\rangle\langle1|$, where $\tau_j=q^j/(1+q)$, $j=0,1$.

The collision between the system qubit and the molecule qubit is described by a two-qubit energy-preserving unitary $U$, which is parametrized as
\begin{equation}
    U=\left(\begin{array}{cccc}
        1 & 0 & 0 & 0 \\
        0 & u_{00} & u_{01} & 0 \\
        0 & u_{10} & u_{11} & 0 \\
        0 & 0 & 0 & e^{i\phi}
    \end{array}\right).
\end{equation}
After $N$ collisions, the state of the qubit becomes $\rho^{(N)}=T^{\circ N}(\rho)$, where $T(\cdot)=\tr_r[U(\cdot\otimes\tau)U^\dagger]$. The result in Ref.~\cite{PhysRevLett.88.097905_collision_model} gives
\begin{eqnarray}
    \rho_{00}^{(N)}&=&\tau_0+|u_{00}|^{2N}(\rho_{00}-\tau_0),\label{eq:rho00}\\
    \rho_{01}^{(N)}&=&[(1-\tau_0)u_{00}e^{-i\phi}+\tau_0u_{11}^*]^N\rho_{01}.\label{eq:rho01}
\end{eqnarray}

Case 1: $z\neq z_\tau$.

From Eq.~(\ref{eq:rho00}), we have
\begin{equation}
    |u_{00}|^N=\sqrt{\frac{z^{(N)}-z_\tau}{z-z_\tau}}.\label{eq:cone_qubit1}
\end{equation}
From Eq.~(\ref{eq:rho01}), we have
\begin{equation}
    \eta^{(N)}=\eta|u_{00}|^N\cdot|(1-\tau_0)+\tau_0 e^{i\alpha}|^N,
\end{equation}
where $\alpha$ is a phase depending on $\phi$ and the phase difference between $u_{00}$ and $u_{11}^*$. Because $|(1-\tau_0)+\tau_0 e^{i\alpha}|\in[2\tau_0-1,1]$ and $z_\tau=2\tau_0-1$, we have 
\begin{equation}
    \eta z^N_\tau |u_{00}|^N\leq\eta^{(N)}\leq\eta|u_{00}|^N.\label{eq:cone_qubit2}
\end{equation}
Substituting Eq. (\ref{eq:cone_qubit1}) to Eq.~(\ref{eq:cone_qubit2}), we arrive at Eq.~(\ref{eq:cone_qubit}).

Case 2: $z=z_\tau$.

From Eq.~(\ref{eq:rho00}), $z^{(N)}=z_\tau$ independent of the value of $u_{00}$. Thus, from Eq.~(\ref{eq:rho01}), $\eta^{(N)}$ can reach any value in the interval $[0,\eta]$ by varying $u_{00}$. The $\mC_\beta(H^{(2)},N)$-cone then consists of states with $z^{(N)}=z_\tau$ and $\eta^{(N)}\in[0,\eta]$. 

This completes the proof of $\mC_\beta(H^{(2)},H^{(2)},N)$-cone for any input qubit states.

It is worth noticing that the $\mC_\beta(H^{(2)},H^{(2)},N)$-cone is nonconvex except for the special cases with $\eta=0$ or $z=z_\tau$. This nonconvexity is caused by the finite size of the environment. When the number of collisions $N\rightarrow\infty$, the $\mC_\beta(H^{(2)},H^{(2)},\infty)$ cone is convex for any input qubit state. Precisely, the necessary and sufficient condition for qubit state transition under $\mC_\beta(H^{(2)},H^{(2)},\infty)$ is
\begin{equation}
    \eta'\in \eta\sqrt{\frac{z'-z_\tau}{z-z_\tau}}[0,1],
\end{equation}
if $z\neq z_\tau$, and $\eta'\in[0,\eta]$ if $z=z_\tau$. This coincides with the necessary and sufficient condition for qubit state transition under MTO derived in \cite{PhysRevA.96.032109}.

Importantly, $\mC_\beta(H^{(2)},H^{(2)},N)$-cone is a subset of the convex hull of $\mC_\beta(H^{(2)},H^{(2)},1)$-cone for any $N$. It means that any extreme point of the convex hull of states achieved by collision model can be realized in one collision. Therefore, if we aim at preparing the extreme points, instead of simulating the time evolution of states, it is adequate to consider only one collision. This dramatically reduces the time cost $N$.

\section{$\beta$-ordering and effective temperature \label{app:beta_ordering}}
Consider a $d$-dimensional quantum system with Hamiltonian $H=\sum_{j=0}^{d-1}E_j|j\rangle\langle j|$. Without loss of generality, $E_j\leq E_{j'}$ for all $0\leq j<j'\leq d-1$. When the system is in a quantum state described by a density matrix $\rho$, the probability $p_j=\langle j|\rho|j\rangle$ is called the population on $|j\rangle$, and  $\vec{p}=\{p_j\}_{j=0}^{d-1}$ is called the energy distribution.

\begin{definition}\cite{Lostaglio_2019}
    ($\beta$-ordering). Let $\vec{p}=\{p_j\}_{j=0}^{d-1}$ be an energy distribution, where $p_j$ is the probability in $|j\rangle$ with energy $E_j$, and $\pi$ be the permutation ensuring
    \begin{equation}
        p_{\pi(0)}e^{\beta E_{\pi(0)}}\geq p_{\pi(1)}e^{\beta E_{\pi(1)}}\geq\dots\geq p_{\pi(d-1)}e^{\beta E_{\pi(d-1)}}.
    \end{equation}
    Then the sequence $\left(\pi(0),\dots,\pi(d-1)\right)$ is called the $\beta$-ordering of $\vec{p}$, and the permutation $\pi$ is said to $\beta$-order $\vec{p}$.
\end{definition}

For example, if the energy distribution $\vec{p}=(p_0,p_1,p_2)$ of a qutrit state satisfies $p_1e^{\beta E_1}\geq p_2e^{\beta E_2}\geq p_0e^{\beta E_0}$, the permutation that $\beta$-orders $\vec{p}$ reads $\pi(0)=1,\pi(1)=2,\pi(2)=0$, and the $\beta$-ordering of $\vec{p}$ is $(1,2,0)$.

If the state of the $d$-dimension system can be written in the thermal form $\tau(H,\beta')$, the inverse temperature of the system is $\beta'$.
The $\beta$-ordering of $\tau(H,\beta')$ is $(0,1,\dots,d-1)$ if $\beta'\geq\beta$, and $(d-1,d-2,\dots,1,0)$ if $\beta'\leq\beta$. Notice the latter case includes states with negative temperature $\beta'<0$. If a system state $\rho$ cannot be written in the thermal form, but $\beta$-ordering of its energy distribution is $(d-1,d-2,\dots,1,0)$, then from Ref.~\cite{PhysRevX.8.041049crude_TO}, there exist $\beta'<\beta$ such that $\tau(H,\beta')\rightarrow\rho$ via partial level thermalization (PLT). Hence, one can define the maximum value of $\beta'$ such that PLT:$\tau(H,\beta')\rightarrow\rho$ as the effective inverse temperature of $\rho$. Similarly, for a state $\rho$ with $\beta$-ordering $(0,1,\dots,d-1)$, its effective temperature is the minimum value of $\beta'$ such that PLT:$\tau(H,\beta')\rightarrow\rho$.

\section{Calculation of $\mC_\beta(H^{(3)},H^{(3)},1)$-cone of qutrit states \label{app:qutrit_cone}}
Before the collision, the system is in the state $\rho$ with the energy distribution $\vec{p}=\{p_0,p_1,p_2\}$ 
and the molecule is in the Gibbs state $\tau^r=\tau(H^{(3)},\beta)$.

The total Hamiltonian can be written as
\begin{equation}
    H^{tot}=E\Pi_1+2E\Pi_2+3E\Pi_3+4E|22\rangle\langle 22|,
\end{equation}
where $\Pi_1=|01\rangle\langle 01|+|10\rangle\langle10|$, $\Pi_2=|02\rangle\langle 02|+|11\rangle\langle11|+|20\rangle\langle20|$, and $\Pi_3=|12\rangle\langle12|+|21\rangle\langle21|$ are projections to the energy eigenspaces of $E$, $2E$, and $3E$, respectively. 
\begin{widetext}
The transition matrices $G_j$ in the eigenspace of $jE$ can be parametrized as
\begin{eqnarray}
    G_1 & = &\left(\begin{array}{cc}
        a_1 & 1-a_1 \\
        1-a_1 & a_1
    \end{array}\right),
    \ \ \ G_3 =\left(\begin{array}{cc}
        a_3 & 1-a_3 \\
        1-a_3 & a_3
    \end{array}\right),\nonumber\\
    G_2 & = &\left(\begin{array}{ccc}
        a_2 & b_2 & 1-a_2-b_2 \\
        a'_2 & b'_2 & 1-a'_2-b'_2 \\
        1-a_2-a'_2 & 1-b_2-b'_2 & a_2 + b_2 + a'_2 + b'_2 -1
    \end{array}\right),    
\end{eqnarray}
where the parameters $a_j,b_j,a'_j,b'_j$ are such that all of the matrix elements in $G_j$ are nonnegative.

The resulted population distribution in the output of the collision is calculated as
\begin{eqnarray}
    p_0^{(1)} & = & \tau^r_0+(a_1+qa_2)(\tau^r_1p_0-\tau^r_0 p_1) + (a_2+b_2)(\tau^r_1p_1-\tau^r_0p_2),\nonumber\\
    p_1^{(1)} & = & \tau^r_1+(a_1-qa'_2)(-\tau^r_1p_0+\tau^r_0 p_1) + (1-a'_2-b'_2-q a_3)(-\tau^r_1p_1+\tau^r_0p_2),\nonumber\\
    p_2^{(1)} & = & \tau^r_2+q(a_2+a'_2)(-\tau^r_1p_0+\tau^r_0 p_1)+(a_2 + b_2 + a'_2 + b'_2 -1+q a_3)(-\tau^r_1p_1+\tau^r_0p_2).\label{eq:qutrit_output}
\end{eqnarray}
\end{widetext}

In order to calculate the $\colli_\beta(H^{(3)},1)$-cone for population dynamics, we first determine the extreme points by calculating the range of $p'_j$, and the other population distribution in the cone can be obtained by continuously varying $a_j,b_j,a'_j,b'_j$. It is easy to see that the range of $p'_j$ differs for different $\beta$-ordering of the input state. Therefore, we divide the set of energy distributions of a $3$-level system into six subsets, each with a fixed $\beta$-ordering:

Subset I: $\beta$-ordering (0,1,2);

Subset II: $\beta$-ordering (2,1,0);

Subset III: $\beta$-ordering (1,0,2);

Subset IV: $\beta$-ordering (2,0,1);

Subset V: $\beta$-ordering (1,2,0);

Subset VI: $\beta$-ordering (0,2,1).

Here we calculate the cone for states initially in subset V. The cone of other initial state can be obtained similarly.

For an initial state in subset V, we have $\tau^r_1p_1\geq\tau^r_0p_2\geq\tau^r_2p_0$. It follows directly from Eq. (\ref{eq:qutrit_output}) that 
\begin{eqnarray}
    p^{(1)}_0 & \in & [p_0,\tau^r_0+(\tau^r_1 p_1-\tau^r_0 p_2)],\nonumber\\
    p^{(1)}_1 & \in & [\tau^r_1-(\tau^r_1p_1-\tau^r_2 p_0),p_1],\nonumber\\
    p^{(1)}_2 & \in & [p_2-(\tau^r_0p_2-\tau^r_2 p_0), \tau^r_2+(\tau^r_1p_1-\tau^r_2p_0)].\label{eq:qutrit_cone}
\end{eqnarray}
In the following, we label the lower bound for $p'_j$ as $(p'_j)_{\min}$ while the upper bound as $(p'_j)_{\max}$, and then the six extreme points of the above region are expressed as
\begin{eqnarray*}
    A_0 & = & [(p_0^{(1)})_{\min},(p_1^{(1)})_{\max},1-(p_0^{(1)})_{\min}-(p_1^{(1)})_{\max}],\\
    A_1 & = & [(p_0^{(1)})_{\min},1-(p_0^{(1)})_{\min}-(p_2^{(1)})_{\max},(p_2^{(1)})_{\max}],\\
    A_2 & = & [1-(p_1^{(1)})_{\min}-(p_2^{(1)})_{\max},(p_1^{(1)})_{\min},(p_2^{(1)})_{\max}],\\
    A_3 & = & [(p_0^{(1)})_{\max},(p_1^{(1)})_{\min},1-(p_1^{(1)})_{\min}-(p_0^{(1)})_{\max}],\\
    A_4 & = & [(p_0^{(1)})_{\max},1-(p^{(1)}_0)_{\max}-(p^{(1)}_2)_{\min},(p_2^{(1)})_{\min}],\\
    A_5 & = & [1-(p^{(1)}_1)_{\max}-(p^{(1)}_2)_{\min},(p_1^{(1)})_{\max},(p_2^{(1)})_{\min}].
\end{eqnarray*}
These extreme points can all be reached by proper $a_j,b_j,a'_j,b'_j$. For example, from Eq. (\ref{eq:qutrit_output}), $(p_0^{(1)})_{\max}=\tau^r_0+(\tau^r_1 p_1-\tau^r_0 p_2)$ is reached when $a_1=a_2=0$ and $b_2=1$, and $(p^{(1)}_2)_{\min}=p_2-(\tau^r_0p_2-\tau^r_2 p_0)$ is reached when $a_2+a'_2=0$, $b_2+b'_2=1$, and $a_3=1$. It follows that the extreme point $A_4=[\tau^r_0+(\tau^r_1 p_1-\tau^r_0 p_2),\tau^r_1(p_0+p_1)+(\tau^r_0p_2-\tau^r_2p_0,p_2-(\tau^r_0p_2-\tau^r_2 p_0)]$ can be reached when
\begin{equation}
    G_1=\left(\begin{array}{cc}
        0 & 1 \\
        1 & 0
    \end{array}\right), 
    G_2=\left(\begin{array}{ccc}
        0 & 1 & 0 \\
        0 & 0 &1 \\
        1 & 0 & 0
    \end{array}\right),
    G_3=\iden_2.
\end{equation}
The transition matrices to reach the other (non-extreme) points in the region described by Eq.~(\ref{eq:qutrit_cone}) can be obtained by solving Eq.~(\ref{eq:qutrit_output}). There are six variables and three equations, so the solutions always exist. Therefore, the $\colli_\beta(H^{(3)},H^{(3)},1)$-cone of any state in Subset V is described in Eq.~(\ref{eq:qutrit_cone}).

\section{Proof of Theorem~\ref{th:no_go}\label{app:no_go}}
The proof of Theorem~\ref{th:no_go} in the main text is as follows.

Before collision, the state of the composed system reads
\begin{equation}
    \rho_{tot}=\rho\otimes\tau^r=\sum_{k=0}^{d_S-1}\sum_{j=0}^{d_r-1}p_k\tau^r_j|kj\rangle\langle kj|.
\end{equation}
Let us denote the state after collision as
\begin{equation}
    \rho^{(1)}_{tot}=\sum_{k=0}^{d_S-1}\sum_{j=0}^{d_r-1}\xi_{kj}|kj\rangle\langle kj|.
\end{equation}
Hence, the ground state population of $S$ in the output reads
\begin{equation}
    p^{(1)}_0=\sum_{j=0}^{d_r-1}\xi_{0j}.
\end{equation}
In order to maximize $p^{(1)}_0$ over all energy-preserving unitaries $U$, the optimal strategy is to reorder the population in each energy subspace such that $\xi_{0j}\geq\xi_{k'j'}$, $\forall k',j'$ satisfying 
\begin{equation}\label{eq:subspace_energy}
  E^S_{k'}+E^r_{j'}=E_0^S+E_j^r.  
\end{equation}
In other words, the optimal strategy gives
\begin{equation}\label{eq:xi_m}
    \xi_{0j}^*=\max_{j',k':E^S_{k'}+E^r_{j'}=E_0^S+E_j^r}p_{k'}\tau^r_{j'}=p_j\tau^r_0.
\end{equation}
The last equality is derived as follows. From Condition (R1) and Eq. (\ref{eq:subspace_energy}), $k'\leq j$. It follows that
\begin{equation}
    p_j\geq p_{k'}e^{-\beta(E_j^S-E_{k'}^S)}=p_{k'}e^{-\beta(E_{j'}^r-E_0^r)},
\end{equation}
where the first inequality is from Condition (R3) and the second equality is from Condition (R2) and Eq. (\ref{eq:subspace_energy}). Equivalently, we have $p_{k'}e^{-\beta E_{j'}^r}\leq p_{j} e^{-\beta E_0^r}$, which in turn gives $p_{k'}\tau_{j'}^r\leq p_j\tau_0^r$ for all $j',k'$ satisfying Eq. (\ref{eq:subspace_energy}).

From Eq. (\ref{eq:xi_m}), we have
\begin{eqnarray}
    p_0^{(1)*}&=&\sum_{j=0}^{d_r-1}p_j\tau^r_0\nonumber\\
    &=&\frac{\sum_{j=0}^{d_r-1}p_j}{\sum_{j=0}^{d_r-1}e^{-\beta E_j^r}}e^{-\beta E_0^r}\nonumber\\
    &=&\frac{\sum_{j=0}^{d_r-1}p_j}{\sum_{j=0}^{d_r-1}e^{-\beta E_j^S}}e^{-\beta E_0^S}\nonumber\\
    &\leq&\frac{\sum_{j=0}^{d_S-1}p_j}{\sum_{j=0}^{d_S-1}e^{-\beta E_j^S}}e^{-\beta E_0^S}=\tau^S_0.
\end{eqnarray}
The equality holds if and only if $\rho=\tau^S$ or $m=d$. This completes the proof of Theorem \ref{th:no_go}.

It should be noticed that, in general, the condition (R2) is essential for Theorem \ref{th:no_go} to hold. In fact, we can find a counterexample in the simplest, non-trivial situation that violates (R2): the Hamiltonians of $S$ and $r$ are respectively $H_S=\sum_{j=0}^{2}jE|j\rangle\langle j|$ and $H_r= 2E|1\rangle\langle 1|$. This means that the common energy gap does not involve the two lowest states of the system ($E_1^S-E_0^S=E$), but rather the ground state and the most excited state ($E_2^S-E_0^S=2E$). Suppose that the state of $S$ is in prepared the most excited state, $p_2=1$. Conditions (R1) and (R3) are satisfied. The non-trivial energy preserving $U$ swaps the populations of states $|2_S0_r\rangle$ and $|0_S1_r\rangle$ and keeps other populations untouched can bring the system to the state with $p_0'=1/(1+q^2)>1/(1+q+q^2)=\tau^S_0$, where $q=e^{-\beta E}$.

\section{Proof of theorem~\ref{th:no_go1}\label{app:no_go1}}
Now the Hilbert space of $S$ can be decomposed as $\mH_S=\mH_{S_1}\otimes\mH_{S_2}$, where $\mH_{S_1}=\mH_r$.
Let $\{|k_\eta\rangle\}_{k_\eta}$ be the energy eigenbasis for the $\eta$th particle, and then, the basis $\{|K\rangle\}_K$ of $H_S$ reads
        \begin{equation}
            |K\rangle=|K_1\rangle\otimes|K_2\rangle,
        \end{equation}
        where $|K_1\rangle=|k_1\dots k_{\nu}\rangle$ and $|K_2\rangle=|k_{\nu+1}\dots k_{\mu}\rangle$.

From (R3), for all $K_1$ and $K_2,K_2'$ satisfying $E_{K_2}^{(S_2)}\geq E_{K_2'}^{(S_2)}$, the following inequality holds
        \begin{equation}
            \frac{p_{K_1K_2}}{e^{-\beta(E_{K_1}^{(S_1)}+E_{K_2}^{(S_2)})}}\geq \frac{p_{K_1K'_2}}{e^{-\beta(E_{K_1}^{(S_1)}+E_{K'_2}^{(S_2)})}},
        \end{equation}
        It follows that
        \begin{equation}
            \frac{\sum_{K_1}p_{K_1K_2}}{\sum_{K_1}e^{-\beta(E_{K_1}^{(S_1)}+E_{K_2}^{(S_2)})}}\geq \frac{\sum_{K_1}p_{K_1K'_2}}{\sum_{K_1}e^{-\beta(E_{K_1}^{(S_1)}+E_{K'_2}^{(S_2)})}},
        \end{equation}
        and hence,
        \begin{equation}
            \frac{p^{(S_2)}_{K_2}}{e^{-\beta E_{K_2}^{(S_2)}}}\geq \frac{p^{(S_2)}_{K'_2}}{e^{-\beta E_{K'_2}^{(S_2)}}},
        \end{equation}
        where $p^{(S_2)}_{K_2}\equiv\sum_{K_1}p_{K_1 K_2}$ is the population of system $S_2$ on state $|K_2\rangle$. It means that for the subsystem $S_2$, the $\beta$-ordering of the initial state is the same as a thermal state at some higher temperature. Therefore, 
        \begin{equation}
            p_\mathbb{O}^{(S_2)}\leq \tau_\mathbb{O}^{(S_2)},\label{eq:s2_p0}
        \end{equation} 
        where $\mathbb{O}$ denotes that all of the particles in the relevant system are in the ground state, and $\tau_\mathbb{O}^{(S_2)}$ is the ground state population of $\tau(H_{S_2},\beta)$.

        Next let us derive the upper bound of the ground state population in the output. Before the collision, the state reads
        \begin{equation}
            \rho_{tot}=\sum_{K_1,K_2}\sum_{J}p_{K_1K_2}\tau_J|K_1K_2J\rangle\langle K_1K_2J|,
        \end{equation}
where $|J\rangle\equiv|j_1\dots j_\nu\rangle$ are energy eigenstates of the molecule $r$. It is important to notice that (R3) implies $p_{K}=p_{K'}$ whenever $E_{K}=E_{K'}$. Hence, $p_{K_1K_2}=p_{K_1'K_2'}$ if $E^{(S_1)}_{K_1}+E^{(S_2)}_{K_2}=E^{(S_1)}_{K'_1}+E^{(S_2)}_{K'_2}$. After the collision, the state can be written as
        \begin{equation}
            \rho^{(1)}_{tot}=\sum_{K_1,K_2}\sum_{J}\xi_{K_1K_2J}|K_1K_2J\rangle\langle K_1K_2J|,
        \end{equation}
        and then, the ground state population for $S$ in the output reads
        \begin{equation}
            p^{(1)}_0=\sum_J\xi_{\mathbb{O}\mathbb{O}J}=\sum_{\mathcal{J}}{\sum_{j_1\dots j_\nu}}^{(\mathcal{J})}\xi_{\mathbb{O}\mathbb{O}J}.
        \end{equation}
        The index $\mathcal{J}$ is for the energy levels $E_\mathcal{J}$ of the molecule, and the summation $\sum_{j_1\dots j_\nu}^{(\mathcal{J})}$ is over indices satisfying $\epsilon_{j_1}+\dots+\epsilon_{j_\nu}=E_{\mathcal{J}}$. From (R3), $p_{J\mathbb{O}}\tau^r_{\mathbb{O}}\geq p_{K'_1K'_2}\tau^r_{J'}$ for all $E_{J}^{(S_1)}=E_{K'_1}^{(S_1)}+E_{K'_2}^{(S_2)}+E_J^{(r)}=E_{\mathcal{J}}$. Hence,
        \begin{equation}
            \xi_{\mathbb{O}\mathbb{O}J}^*=p_{J\mathbb{O}}\tau^r_{\mathbb{O}}.
        \end{equation}
        It follows that
        \begin{eqnarray}
            p_0^{(1)*}&=&\sum_{\mathcal{J}}{\sum_{j_1\dots j_\nu}}^{(\mathcal{J})}p_{J\mathbb{O}}\tau^r_{\mathbb{O}}\nonumber\\
            &=& p_{\mathbb{O}}^{(S_2)}\tau^r_{\mathbb{O}}\nonumber\\
            &\leq& \tau_{\mathbb{O}}^{(S_2)}\tau_{\mathbb{O}}^{(S_1)}=\tau^S_{\mathbb{O}}.
        \end{eqnarray}
        The reason for the second line is that the Hilbert spaces for $S_1$ and $r$ are the same, and the third line is from Eq.~(\ref{eq:s2_p0}) and $\tau^r_{\mathbb{O}}=\tau_{\mathbb{O}}^{(S_1)}$. The equation in the third line holds if and only if $p_{\mathbb{O}}=\tau^S_{\mathbb{O}}$ or $S_2=\emptyset$. This completes the proof.

\begin{widetext}
\section{Protocols in the absence of a machine \label{app:protocol}}
By constructive proof, we show that the ground state population in the output converges to
\begin{equation}
    p_0^*=\left\{\begin{array}{ll}
    \frac{1-q^{d_r-1}}{1-(q^{d_r-1})^{d_S}}, & d_r=3,\\
    \frac{1-q^{d_r+2}}{(1+q^{d_r-1})\cdot[1-(q^{d_r+2})^k]}, & d_r\geq 4,d_S=2k,\\
    \left[\frac{(1+q^{d_r-1})\cdot[1-(q^{d_r+2})^k]}{1-q^{d_r+2}}+q^{(d_r+2)k-1}\right]^{-1}, & d_r\geq4,d_S=2k+1.
        \end{array}\right.
\end{equation}

\subsection{Cooling protocol with qutrit molecules}
In this case, we have $d_r=3$ and $d_S\geq3$. The total Hamiltonian of $S$ and $r$ is
\begin{equation}
    H_{Sr}=\sum_{j=1}^{d_S}jE\Pi_j+(d_S+1)E|d_S-1,2\rangle\langle d_S-1,2|,
\end{equation}
where $\Pi_j$ is the projection onto the eigen subspace of $jE$. In the following, we will specify the explicit form of $V$ in the recharging routine $\mathcal{V}(\cdot)=V(\cdot)V^\dagger$, as well as that of the global unitary $U$ in the thermalizing routine $\Gamma(\cdot)=\tr_r[U(\cdot\otimes\tau^r)U^\dagger]$, such that the following transition matrix $G_{\mE\circ \mathcal{V}}=G^*$ where
\begin{eqnarray}
   & G^*_{00}=\tau^r_0+\tau^r_1, \ G^*_{01}=\tau^r_0, \nonumber\\
   & G^*_{j,j-1}=\tau^r_2, \ G^*_{j,j}=\tau^r_1, G^*_{j,j+1}=\tau^r_0,\  j=1,\dots,d_S-2,\nonumber\\
   & G^*_{d_S-1,d_S-2}=\tau^r_2, \ G^*_{d_S-1,d_S-1}=\tau^r_1+\tau^r_2,\label{eq:eff_G_S}
\end{eqnarray}
can be achieved.
The unique fixed point of $G^*$ reads
\begin{equation}
    \vec{p}_*=\frac{1}{\mathcal{N}}(1,q^2,\dots,q^{2(d_S-1)})^\mathrm{T}.\label{eq:fix_point}
\end{equation}
It is a Gibbs state of the $d_S$-dimensional target system at temperature $\beta'=2\beta$.

For latter convenience, we label
\begin{equation}
    u=\left(\begin{array}{ccc}
        1 & 0 & 0 \\
        0 & 0 & 1 \\
        0 & 1 & 0
    \end{array}\right),
     G_a =\left(\begin{array}{ccc}
        0 & 0 & 1 \\
        1 & 0 & 0 \\
        0 & 1 & 0
    \end{array}\right),G_b =\left(\begin{array}{ccc}
        0 & 1 & 0 \\
        0 & 0 & 1 \\
        1 & 0 & 0
    \end{array}\right).
\end{equation}
Besides, $G_j$ denotes the transition matrix induced by $U$ in the energy subspace of $H_{Sr}$ associated with $jE$.

For $d_S=3k+1$, in order to achieve the effective transition matrix described by Eq. (\ref{eq:eff_G_S}), we set
\begin{equation}
    V=\underbrace{ u\oplus u\dots \oplus u}_k\oplus  (1),
\end{equation}
where $(1)$ stands for a $1\times 1$ matrix with element $1$. In other words, the unitary $V$ exchanges the populations on $|3j-1\rangle$ and $|3j\rangle$ for $j=1,\dots,k$. The transition matrices in each degenerate subspace induced by $U$ is
\begin{eqnarray}
    G_1  =  \iden_2, & G_2=G_a, & G_3 = G_b,\nonumber\\
    G_{3j+1}=\iden_3, & G_{3j+2}=G_a, & G_{3j+3}=G_b,j=1,\dots, k-1,\nonumber\\
    & G_{3k+1}=\iden.  &
\end{eqnarray}

For $d_S=3k+2$, we have
\begin{equation}
    V=\sigma_x\oplus\underbrace{ u\oplus u\dots \oplus u}_k,
\end{equation}
and 
\begin{eqnarray}
    G_1=G_{3k+2}=\sigma_x,\nonumber\\
    G_{3j}=\iden_3, 
    G_{3j-1}=G_b, G_{3j+1}=G_a,
\end{eqnarray}
where $j=1,\dots,k$.

For $d_S=3k$, we have
\begin{equation}
    V=\underbrace{ u\oplus u\dots \oplus u}_k,
\end{equation}
and
\begin{eqnarray}
    G_1=\iden, G_2=G_a,G_{3k}=\sigma_x,\nonumber\\
    G_{3j}=G_b, G_{3j+1}=\iden, G_{3j+2}=G_a, 
\end{eqnarray}
where $j=1,\dots,k-1$.

For all three cases above, the transition matrix described by Eq. (\ref{eq:eff_G_S}) can be achieved.

\subsection{$d_r\geq 4$}
The state of the target can converge to $\vec{p}_*$.

For $d_S=2k$, $k\geq 2$, the elements of $\vec{p}_*$ satisfy
\begin{eqnarray}
    p^*_{2j-1}=q^{d_r-1}p^*_{2j-2},\ j=1,\dots,k,\nonumber\\
    p^*_{2j}=q^3p^*_{2j-1},\ j=1,\dots,k-1.\label{eq:fix_poi_2k_d_geq_4}
\end{eqnarray}
Consequently, we have
\begin{equation}
    p^*_l=\left\{\begin{array}{cc}
        \frac{1}{Z_{2k}^*}(q^{d_r+2})^{j-1}, & l=2j-2, \\
        \frac{1}{Z_{2k}^*}q^{d_r-1}(q^{d_r+2})^{j-1}, & l=2j-1,
    \end{array}\right.
\end{equation}
where $j=1,\dots,k$, and
\begin{equation}
    Z_{2k}^*=(1+q^{d_r-1})\cdot\frac{1-(q^{d_r+2})^k}{1-q^{d_r+2}}.
\end{equation}

For $d_S=2k+1$, $k\geq 2$, the elements of $\vec{p}_*$ satisfy
\begin{eqnarray}
    p^*_{2j-1}=q^{d_r-1}p^*_{2j-2},\ p^*_{2j}=q^3p^*_{2j-1},\ \nonumber\\
    p^*_{2k-1}=q^{d_r-1}p^*_{2k-2},\ p^*_{2k}=q^2p^*_{2k-1}.
    \label{eq:fix_poi_2kp1_d_geq_4}
\end{eqnarray}
where $j=1,\dots,k-1$.
Consequently, we have
\begin{equation}
    p^*_l=\left\{\begin{array}{cc}
        \frac{1}{Z_{2k+1}^*}(q^{d_r+2})^{j-1}, & l=2j-2, \\
        \frac{1}{Z_{2k+1}^*}q^{d_r-1}(q^{d_r+2})^{j-1}, & l=2j-1,\\
        \frac{1}{Z_{2k+1}^*}q^{(d_r+2)k-1}, & l=2k,
    \end{array}\right.
\end{equation}
where $j=1,\dots,k$, and
\begin{equation}
    Z_{2k+1}^*=(1+q^{d_r-1})\cdot\frac{1-(q^{d_r+2})^k}{1-q^{d_r+2}}+q^{(d_r+2)k-1}.
\end{equation}

A few discussions are in order.

Firstly, for $d_S=2k$ and $d_r=4$, Eq. (\ref{eq:fix_poi_2k_d_geq_4}) reduces to $p_{l+1}=q^{3}p_l$ for $l=0,\dots,d_S-2$. It means that $\beta'=(d_r-1)\beta$ can be reached. For other cases, the asymptotic state deviates from the Gibbs state $\tau(H_S,(d_r-1)\beta)$. However, it is noticed that $p^*_1=q^{d_r-1}p^*_0$ for any $d_S$. When the temperature is low or energy gap is large such that $q\ll 1$, the populations on higher energy level are negligible. Hence, the asymptotic state is close to $\tau(H_S,(d_r-1)\beta)$, i.e., an effective temperature $\beta'=(d_r-1)\beta$ can be approximately approached.

Secondly, in the recharging routine, only population exchanges between adjacent energy levels are employed (see Eqs. (\ref{eq:v_2k}) and (\ref{eq:v_2k1}) below). The energy consumed in $n$-th cycle is calculated as
\begin{eqnarray}
    W^{(n)}&=&E\sum_{j=0}^{k-2}(p^{(n)}_{2j}-p^{(n)}_{2j+1})\nonumber\\
    &\leq&E\sum_{j=0}^{k-2}(p^{(\infty)}_{2j}-p^{(\infty)}_{2j+1})\nonumber\\
    &=&E(1-q^{d_r-1})\sum_{j=0}^{k-2}p^{(\infty)}_{2j}\nonumber\\
    &=&E(1-q^{d_r-1})p_0^{(\infty)}\sum_{j=0}^{k-2}(q^{d_r+2})^j\nonumber\\
    &\leq&E\cdot\frac{(1-q^{d_r-1})\sum_{j=0}^{k-2}(q^{d_r+2})^j}{(1+q^{d_r-1})\sum_{j=0}^{k-2}(q^{d_r+2})^j}\nonumber\\
    &=&E\cdot\frac{(1-q^{d_r-1})}{(1+q^{d_r-1})}\leq E.
\end{eqnarray}
As for xHBAC under coherent control, the recharging routine transforms the target state to the most activated state, exchanging the populations on $|j\rangle$ and $|d_S-1-j\rangle$, where $j=0,\dots,d_S-1$. The consumed energy in each round is lower bounded by that consumed in the first round calculated as
\begin{eqnarray}
    W_{\mathrm{xHBAC}}^{(1)}&=&E\sum_{j=0}^{d_S-1}(d_S-1-2j)\tau^S_{j}\nonumber\\
    &=&(d_S-1)E-2E\sum_{j=0}^{d_S-1}j\tilde\tau^r_{j}\nonumber\\
    &=&(d_S-1+\frac{2d_Sq}{1-q^{d_S}}-\frac{2q}{1-q})E,
\end{eqnarray}
which diverges for large $d_S$.
    
The motivation in obtaining the protocol is as follows. Assume that $\tau(H_S,(d_r-1)\beta)$ can be approached by our cooling protocol where the operation in the thermalizing routine belongs to $\mC(H^{(d_r)},H^{(d_S)},1)$, and then the transition matrix $\bar{G}_S$ on $S$ of a single round should satisfy the following two conditions:

(i) the non-zero elements of $\bar{G}_S$ should be summation of a subset of $\{\tau^r_j\}_{j=0,\dots,d_r-1}$.

(ii) the fixed point of $\bar{G}_S$ should be
\begin{equation}
    \vec{p}_{**}=\left\{\frac{q^{k(d_r-1)}}{Z_{d_S}((d_r-1)\beta)}\right\}_{k=0,\dots,d_S-1}.
\end{equation}

The only solution of $\bar{G}_S$ which satisfies these two conditions is in the following form
\begin{equation}
    \bar{G}_S=\left(\begin{array}{cccccc}
        1-\tau^r_{d_r-1} & \tau^r_0 & 0 & \cdots & 0 & 0 \\
        \tau^r_{d_r-1} & 1-\tau^r_0-\tau^r_{d_r-1} & \tau^r_0 & \cdots & 0 & 0\\
        0 & \tau^r_{d_r-1} & 1-\tau^r_0-\tau^r_{d_r-1} & \cdots & 0 & 0 \\
        \vdots & \vdots & \vdots & \ddots & \vdots & \vdots \\
        0 & 0 & 0 & \cdots & 1-\tau^r_0-\tau^r_{d_r-1} & \tau^r_0 \\
        0 & 0 & 0 & \cdots & \tau^r_{d_r-1} & 1-\tau^r_0
    \end{array}\right).
\end{equation}
However, for $d_r\geq5$, $\bar{G}_S\circ G_{\mathcal{V}}$ is not Gibbs-preserving for any bi-stochastic matrix $G_{\mathcal{V}}$. It means that $\bar{G}_S$ cannot be reached by $\mE\circ \mathcal{V}$ for any Gibbs-preserving $\mE$ and any unitary $V$.

Now we relax condition (ii) to be \\
(ii') the fixed point $\vec{p}_*$ of $G_S$ should satisfy
\begin{equation}
    p^*_1=q^{d_r-1}p^*_0,
\end{equation}
and further require that\\
(iii) $G_S$ can be realized by $\mE\circ V$ for some Gibbs-preserving $\mE$ and unitary $V$.

The solution to (i), (ii'), and (iii) is $G^*_S$ in the following form
\begin{equation}
    G^*_S=\left(\begin{array}{cccccc}
        1-\tau^r_{d_r-1} & \tau^r_0 & 0 & \cdots & 0 & 0 \\
        \tau^r_{d_r-1} & 1-\tau^r_0-\tau^r_{d_r-1} & \tau^r_{d_r-4} & \cdots & 0 & 0\\
        0 & \tau^r_{d_r-1} & 1-\tau^r_{d_r-4}-\tau^r_{d_r-1} & \cdots & 0 & 0 \\
        \vdots & \vdots & \vdots & \ddots & \vdots & \vdots \\
        0 & 0 & 0 & \cdots & 1-\tau^r_{d_r-4}-\tau^r_{d_r-1} & \tau^r_0 \\
        0 & 0 & 0 & \cdots & \tau^r_{d_r-1} & 1-\tau^r_0
    \end{array}\right),\ d_S=2k,
\end{equation}
and
\begin{equation}
    G^*_S=\left(\begin{array}{ccccccc}
        1-\tau^r_{d_r-1} & \tau^r_0 & 0 & \cdots & 0 & 0 & 0\\
        \tau^r_{d_r-1} & 1-\tau^r_0-\tau^r_{d_r-1} & \tau^r_{d_r-4} & \cdots & 0 & 0 &0\\
        0 & \tau^r_{d_r-1} & 1-\tau^r_{d_r-4}-\tau^r_{d_r-1} & \cdots & 0 & 0 & 0\\
        \vdots & \vdots & \vdots & \ddots & \vdots & \vdots & \vdots\\
        0 & 0 & 0 & \cdots & 1-\tau^r_{d_r-4}-\tau^r_{d_r-1} & \tau^r_0  & 0\\
        0 & 0 & 0 & \cdots & \tau^r_{d_r-1} & 1-\tau^r_0-\tau^r_{d_r-1} & \tau^r_{d_r-3}\\
        0 & 0 & 0 & \cdots & 0 & \tau^r_{d_r-1} & 1-\tau^r_{d_r-3}
    \end{array}\right),\ d_S=2k+1.
\end{equation}
The fixed point of $G^*_S$ is in the form of Eqs. (\ref{eq:fix_poi_2k_d_geq_4}) and (\ref{eq:fix_poi_2kp1_d_geq_4}).

The protocol for realizing $G^*_S$ is as follows.

For $d_S=2k$, the pumping unitary is
\begin{equation}
    V=\underbrace{\sigma_x\oplus\dots\oplus\sigma_x}_{k}.\label{eq:v_2k}
\end{equation}
The joint unitary permutes the population in each eigen subspace of $H_{Sr}$ such that the output distribution $\vec{p}'$ of $S$ is related to the input distribution $\vec{p}$ as follows
\begin{eqnarray}
    p'_0&=&p_0\tau^r_0+p_1(\tau^r_0+\dots+\tau^r_{d_r-2}),\nonumber\\
    p'_{2j-1}&=&p_{2j-2}(\tau^r_1+\dots+\tau^r_{d_r-2})+p_{2j+1}\tau^r_{d_r-4}+p_{2j-1}\tau^r_{d_r-1},\nonumber\\
    p'_{2j}&=&p_{2j}\tau^r_0+p_{2j+1}(\tau^r_0+\dots+\tau^r_{d_r-5})+p_{2j-2}\tau^r_{d_r-1}+p_{2j+1}(\tau^r_{d_r-3}+\tau^r_{d_r-2}),\nonumber\\
    p_{2k-1}&=&p_{2k-2}(\tau^r_1+\dots+\tau^r_{d_r-1})+p_{2k-1}\tau^r_{d_r-1}.
\end{eqnarray}
where $j=1,\dots,k-1$. The transition matrix in each subspace can be derived from the above expression.

For $d_S=2k+1$, the pumping unitary is
\begin{equation}
    V=\underbrace{\sigma_x\oplus\dots\oplus\sigma_x}_{k}\oplus(1).\label{eq:v_2k1}
\end{equation}
The joint unitary perturbs the population in each degenerate subspace such that the output distribution $\vec{p}'$ of the target is related to the input distribution $\vec{p}$ as follows
\begin{eqnarray}
    p'_0&=&p_0\tau^r_0+p_1(\tau^r_0+\dots+\tau^r_{d_r-2}),\nonumber\\
    p'_{2j-1}&=&p_{2j-2}(\tau^r_1+\dots+\tau^r_{d_r-2})+p_{2j+1}\tau^r_{d_r-4}+p_{2j-1}\tau^r_{d_r-1},\nonumber\\
    p'_{2j}&=&p_{2j}\tau^r_0+p_{2j+1}(\tau^r_0+\dots+\tau^r_{d_r-5})+p_{2j-2}\tau^r_{d_r-1}+p_{2j+1}(\tau^r_{d_r-3}+\tau^r_{d_r-2}),\nonumber\\
    p_{2k-1}&=&p_{2k-2}(\tau^r_1+\dots+\tau^r_{d_r-2})+p_{2k-1}\tau^r_{d_r-1}+p_{2k}\tau^r_{d_r-3},\nonumber\\
    p_{2k}&=&p_{2k-2}\tau^r_{d_r-1}+p_{2k}(\tau^r_0+\dots+\tau^r_{d_r-4}+\tau^r_{d_r-2}+\tau^r_{d_r-1}).
\end{eqnarray}
where $j=1,\dots,k-1$.

\end{widetext}

\section{Protocols with a qubit machine \label{app:w_machine}}
Our setup is the same as the one in the main text. The system and the molecule are both qutrits with Hamiltonian $H^{(3)}=E|1\rangle\langle1|+2E|2\rangle\langle2|$, and the machine is a qubit with Hamiltonian $H_M=E|1\rangle\langle1|$. The initial state of the three systems are all Gibbs states at inverse temperature $\beta$. Protocol I refers to the protocol without a machine, and Protocol II refers to the one with the qubit machine described above.

\subsection{Single-Round Scenario}
In the single-round scenario, the maximum ground state population of $S$ that can be achieved by Protocol I in the output reads
\begin{equation}
\label{eq:gsp_3}
    [p^{(1)}_{\mathrm{I}}]^*=\tau^r_0+\tau^r_0(\tau^r_1-\tau^r_2).
\end{equation}
Again, the superscript $[\cdot]^{(1)}$ stands for the first round, and $[\cdot]^*$ means the reachable optimal ground state population.
The pumping unitary achieving this is $V_{\mathrm{I}}=\sigma_x\oplus(1)$. It follows that the consumed energy for Protocol I is
\begin{equation}
    W_{\mathrm{I}}^{(1)}=qE(\tau^r_0-\tau^r_1).
\end{equation}

For Protocol II, the ground state population in Eq. (\ref{eq:gsp_3}) can be reached by choosing $V_{\mathrm{II}}$ and $\Gamma_{\mathrm{II}}$ defined as follows. The effect of $V_{\mathrm{II}}$ is to exchange the populations on $|01\rangle_{SM}$ and $|11\rangle_{SM}$ while preserve the populations on other states. The effect of $U_{\mathrm{II}}$ in $\Gamma_{\mathrm{II}}$ is to sort the populations within every energy subspace of $H_{SMr}$ in decreasing order. The work consumed in realizing $V_{\mathrm{II}}$ is
\begin{equation}
    W_{\mathrm{II}}^{(1)}=E(\tau^r_0-\tau^r_1)\cdot\frac{q}{1+q},
\end{equation}
which is strictly less than the energy consumed in Protocol I.

It is also interesting to notice that, the maximum ground state population achieved in Protocol II does not outperform that in Protocol I, i.e., $[p^{(1)}_{\mathrm{I}}]^*=[p^{(1)}_{\mathrm{II}}]^*$.

\emph{proof}.
    The effect of $V_{\mathrm{II}}$ and $U_{\mathrm{II}}$ is to reorder the populations, and initially, the qutrit system $S$, the qubit machine $M$ and the molecule $r$ are in the Gibbs state, so the ground state population of $S$ in the output should be in the following form
    \begin{equation}
        p_{\mathrm{II}}^{(1)}=\frac{n_0+n_1q+n_2q^2+n_3q^3}{(1+q)(1+q+q^2)^2},
    \end{equation}
    where $n_j$ are integers satisfying $\sum_{j=0}^3n_j=6$, and are determined by $V_{\mathrm{II}}$ and $U_{\mathrm{II}}$ and independent of $q$. Because the largest energy gap in $Mr$ is $3E$, then from the bound derived in Ref. \cite{PhysRevLett.123.170605},
    \begin{equation}
        p_{\mathrm{II}}^{(1)}\leq\frac{1}{1+q^3+q^6},
    \end{equation}
    which gives
    \begin{equation}
        n_0+n_1q+n_2q^2+o(q^3)\leq 1+3q+5q^2+o(q^3),\forall q,
    \end{equation}
    and in turn,
    \begin{equation}
        n_0\leq 1,\ n_1\leq 3.
    \end{equation}
    Note that $[p^{(1)}_{\mathrm{I}}]^*$ in Eq. (\ref{eq:gsp_3}) can also be expressed as
    \begin{equation}
        [p^{(1)}_{\mathrm{I}}]^*=\frac{1+3q+2q^2}{(1+q)(1+q+q^2)^2}.
    \end{equation}
    Therefore, we have $p_{\mathrm{II}}^{(1)}\leq [p^{(1)}_{\mathrm{I}}]^*$ for any choice of $V_{\mathrm{II}}$ and $U_{\mathrm{II}}$. But by definition, we have $[p^{(1)}_{\mathrm{I}}]^*\leq[p^{(1)}_{\mathrm{II}}]^*$. Hence, $[p^{(1)}_{\mathrm{I}}]^*=[p^{(1)}_{\mathrm{II}}]^*$.

\subsection{Details in the protocol which improved the energy efficiency}

The effective transition matrix on $SM$ in one round is calculated as
\begin{equation}
    G_{\Gamma_{\mathrm{II}}\circ \mathcal V_{\mathrm{II}}} =
    \left(\begin{array}{cccccc}
        0 & \tau^r_0+\tau^r_1 & \tau^r_0 & 0 & 0 & 0 \\
        1 & 0 & 0 & 0 & 0 & 0 \\
        0 & 0 & 0 & \tau^r_0+\tau^r_1 & \tau^r_0 & 0 \\
        0 & \tau^r_2 & \tau^r_1+\tau^r_2 & 0 & 0 & 0 \\
        0 & 0 & 0 & 0 & 0 & 1 \\
        0 & 0 & 0 & \tau^r_2 & \tau^r_1+\tau^r_2 & 0
    \end{array}
    \right).
\end{equation}
The fixed point of the above transition matrix is $\tau(H_S,2\beta)\otimes\frac{\iden}{2}$. 
However, detailed analysis shows that, instead of converging to the fixed point monotonically, the state evolves in an oscillating way. Precisely, it is checked that $G_{\Gamma_{\mathrm{II}}\circ \mathcal V_{\mathrm{II}}}^2$ is block diagonal in the following way.
In the subspace $\mathcal{S}_1$ panned by $\{|00\rangle,|11\rangle,|20\rangle\}$,
\begin{equation}
    [G_{\Gamma_{\mathrm{II}}\circ \mathcal V_{\mathrm{II}}}^2]_{\mathcal{S}_1}=\left(\begin{array}{ccc}
        \tau^r_0+\tau^r_1 & \tau^r_0(\tau^r_0+\tau^r_1) & (\tau^r_0)^2 \\
    \tau^r_2     & (\tau^r_0+\tau^r_1)(\tau^r_1+\tau^r_2) & \tau^r_0(\tau^r_1+\tau^r_2) \\
    0 & \tau^r_2 & \tau^r_1+\tau^r_2
    \end{array}\right).
\end{equation}
In the subspace $\mathcal{S}_2$ panned by $\{|01\rangle,|10\rangle,|21\rangle\}$,
\begin{equation}
    [G_{\Gamma_{\mathrm{II}}\circ \mathcal V_{\mathrm{II}}}^2]_{\mathcal{S}_2}=\left(\begin{array}{ccc}
        \tau^r_0+\tau^r_1 & \tau^r_0 & 0 \\
    \tau^r_2(\tau^r_0+\tau^r_1)     & (\tau^r_0+\tau^r_1)(\tau^r_1+\tau^r_2) & \tau^r_0 \\
    (\tau^r_2)^2 & \tau^r_2(\tau^r_1+\tau^r_2) & \tau^r_1+\tau^r_2
    \end{array}\right).
\end{equation}
It can be checked that, for both $G=[G_{\Gamma_{\mathrm{II}}\circ \mathcal V_{\mathrm{II}}}^2]_{\mathcal{S}_{1,2}}$,
\begin{equation}
    \lim_{k\rightarrow\infty} G^k=\left(\begin{array}{c}
        p_0^* \\
        p_1^* \\
        p_2^* \\
    \end{array}\right)\left(\begin{array}{ccc}
        1 & 1 & 1
    \end{array}\right),
\end{equation}
where $\vec{p}^*=\vec{\tau}(H_S,2\beta)$. Therefore, when the initial distribution is $\vec{p}_{SM}=\tau(H_S,\beta)\otimes\tau(H_M,\beta)=(p_{ij})_{i=0,1,2;j=0,1}$, after $2k$ rounds of operation, the final distribution converges to
\begin{equation}
    (p_{00}+p_{11}+p_{20})\vec{p}^*_{\mathcal{S}_1}\oplus (p_{01}+p_{10}+p_{21})\vec{p}^*_{\mathcal{S}_1}.
\end{equation}
But after $2k+1$ rounds of operation, the final distribution converges to
\begin{equation}
    (p'_{00}+p'_{11}+p'_{20})\vec{p}^*_{\mathcal{S}_1}\oplus (p'_{01}+p'_{10}+p'_{21})\vec{p}^*_{\mathcal{S}_1},
\end{equation}
where $\vec{p}'_{SM}=G_{\Gamma_{\mathrm{II}}\circ \mathcal V_{\mathrm{II}}}\vec{p}_{SM}$. Therefore, the asymptotic distribution of the composed system for even number of rounds is different from that for odd number of rounds.  This oscillation behavior of state dynamics leads to the oscillation of consumed work in each round. 

\subsection{Details in the protocol which improves the cooling limit}
The total Hamiltonian of the composed system $SMr$ is
\begin{equation}
    H_{SMr}=\sum_{j=1}^4 jE\Pi_j+5E|212\rangle\langle212|,
\end{equation}
where $\Pi_j$ is the projection to the eigenspace of $jE$. 
Here the basis of each energy eigenspace is ordered increasingly. For example, the basis of the energy eigenspace associated with $2E$ is $\{|002\rangle,|011\rangle,|101\rangle,|110\rangle,|200\rangle\}$. Let $G_j$ be the transition matrix in the eigenspace of $jE$ induced by $U$. The expression of $G_j$ in the protocol in Sec. \ref{subsub:imp_cooling_limit} is
\begin{eqnarray}
    G_1&=&\left(\begin{array}{ccc}
        0 & 1 & 0 \\
        1 & 0 & 0 \\
        0 & 0 & 1
    \end{array}\right),\ 
    G_4=\left(\begin{array}{ccc}
        1 & 0 & 0 \\
        0 & 0 & 1 \\
        0 & 1 & 0
    \end{array}\right),\nonumber\\
    G_2&=&\left(\begin{array}{ccccc}
        0 & 1 & 0 & 0 & 0 \\
        0 & 0 & 0 & 1 & 0 \\
        1 & 0 & 0 & 0 & 0 \\
        0 & 0 & 1 & 0 & 0 \\
        0 & 0 & 0 & 0 & 1
    \end{array}\right),\nonumber\\
    G_3&=&\left(\begin{array}{ccccc}
        1 & 0 & 0 & 0 & 0 \\
        0 & 0 & 1 & 0 & 0 \\
        0 & 0 & 0 & 0 & 1 \\
        0 & 1 & 0 & 0 & 0 \\
        0 & 0 & 0 & 1 & 0 \\
    \end{array}\right).
\end{eqnarray}

\newpage


\bibliography{apssamp}

\end{document}